\documentclass[a4paper,fleqn,usenatbib]{mnras}

\usepackage{newtxtext,newtxmath}

\usepackage[utf8]{inputenc}
\usepackage{mathptmx}

\usepackage{animate}

\usepackage[T1]{fontenc}
\usepackage{ae,aecompl}

\usepackage{graphicx}	

\title[galaxy groups]{Understanding `galaxy groups' as a unique structure in the universe}

\author[S. Paul et al.]{
S. Paul,$^{1}$\thanks{E-mail: surajit@physics.unipune.ac.in (SP)}
R. S. John,$^{2}$
P. Gupta$^{1}$
and H. Kumar$^{2}$
\\
$^{1}$Department of Physics, SP Pune University, Ganeshkhind, Pune, India 411007\\
$^{2}$Dept. of Physics, Pondicherry Engineering College, Puducherry, 605014, India
}

\date{Accepted XXX. Received YYY; in original form ZZZ}

\pubyear{2015}

\begin{document}
\label{firstpage}
\pagerange{\pageref{firstpage}--\pageref{lastpage}}
\maketitle

\begin{abstract}

`Galaxy groups' have hardly been realised as a separate class of objects with specific characteristics in the structural hierarchy. The presumption that the self-similarity of dark matter structures is a valid prescription for the baryonic universe at all scales has rendered smaller structures undetectable by current observational facilities, leading to lesser dedicated studies on them. Some recent reports that indicate a deviation from $\rm{L_x}$-T scaling in groups compared to clusters have motivated us to study their physical properties in depth. In this article, we report the extensive study on physical properties of groups in comparison to the clusters through cosmological hydrodynamic plus N-body simulations using ENZO 2.2 code. As additional physics, radiative cooling, heating due to supernova and star motions, star formation and stellar feedback has been implemented. We have produced a mock sample of 362 objects with mass ranging from $5\times10^{12}\; \rm{M_{\odot}}$ to 2.5$\times 10^{15}\; \rm{M_{\odot}}$. Strikingly, we have found that objects with mass below $\sim$ $8\times 10^{13}\;\rm{M_{\odot}}$ do not follow any of the cluster self-similar laws in hydrostatics, not even in thermal and non-thermal energies. Two distinct scaling laws are observed to be followed with breaks at $\sim$ $8\times 10^{13}\;\rm{M_{\odot}}$ for mass, $\sim$1 keV for temperature and $\sim$1 Mpc for radius. This places groups as a distinct entity in the hierarchical structures, well demarcated from clusters. This study reveals that groups are mostly far away from virialization, suggesting the need for formulating new models for deciphering their physical parameters. They are also shown to have high turbulence and more non-thermal energy stored, indicating better visibility in the non-thermal regime. 

\end{abstract}




\section{Introduction}\label{intro}
Hierarchical structure formation models predict that the constituents of large scale structures (LSS) should follow the self-similarity \citep{kaiser1986MNRAS,morandi2016MNRAS}. Though, extensive studies on galaxies and galaxy clusters have established their distinct characteristics, intermediate structures are not studied well and for long, they were speculated to be just a scaled down version of the clusters \citep{kaiser1986MNRAS,Vikhlinin2006ApJ}. But, a handful of recent observations and simulations indicate a discrepancy in energy and mass scaling in low mass systems (\cite{Gaspari2011MNRAS,Dave_2002ApJ,Bharadwaj2015,Planelles_2013MNRAS,Paul_2015fers.confE} etc.). In the structural hierarchy, there are intermediate objects with distinct physical properties, possibly loose groups or cabal, that can help us to understand the evolutionary trails of galaxy clusters from galaxies. While, clusters are the most massive bound structures, groups are more numerous and are home to a significant fraction of the entire galaxy population in the universe \citep{Mulchaey2000ARA&A,Kamatsu2002MNRAS}. Unlike clusters that emerges at the filamentary nodes, groups are also form inside the dark matter (DM) filaments connecting the clusters \citep{lietzen2012,tempel2014a}. 

Groups are supposed to be virialized as their sound crossing times are much less than the Hubble time (0.1 - 0.5 $\rm{H_0^{-1}}$). But, in reality, groups are mostly found to be in non-virialized state \citep{Diaferio1993AJ,Diaz2010MNRAS}. Moreover, groups being inside the shallower gravitational potential of filaments, are expected to be strongly affected by mergers, shock heating, feedback from supernova and super-massive black holes, and galactic winds etc. \citep{Lovisari2015}. Mach number dependent cosmic ray content is shown to be larger by almost an order of magnitude in the groups than in the clusters \citep{Jubelgas2008A&A}. Fractional presence and the activity of AGN are also much higher in low-mass systems \citep{Gilmour2007MNRAS,Sivakoff2008ApJ}. Thus, group environment is significantly different than the clusters, leading to the expectation of having different physical properties. This also indicates that many of the scaling laws derived from the cluster properties would not be applicable for groups. Some of the recent researches by \citealp{Sun2012NJPh,Stott_2012MNRAS,Dave_2002ApJ,Gaspari2011MNRAS,Bharadwaj2015,Planelles_2013MNRAS} etc. have indicated such deviations. But, whom should we call as group? No universal definition or classifying characteristics are provided for the groups yet. Also, the above mentioned studies are mostly dealing with either a specific subset of data or are done with the radius and mass at over density ratio of 500. But to verify the self similar scaling, it will be physically more appropriate to work with unbiased complete set of data as well as verify the scaling laws with virialized objects \citep{kaiser1986MNRAS,Cole_1996MNRAS,Miniati_2015Natur} i.e. properties within over-density $\sim 200$, derived from the spherical collapse and dynamical equilibrium model (well known value is 178). So, our study will mostly focus on over-density of 200 to find out a generalised definition or classifying characteristics for ``galaxy groups"  and ``galaxy clusters".

In this study, we intend to understand the distinct features of low mass objects in comparison to the clusters through cosmological hydrodynamic plus N-body simulations. We have theoretically modelled LSS, taking into account the effect of cooling (radiative processes, e.g. X-ray emission), heating (Supernova (SN) and star motions), star formation and star formation feedback physics that we will further call as `coolSF' simulations. AGN feedback has not been considered in this present work which may fine tune our results, mostly, at very low masses i.e. few times of $10^{13} \rm{M_\odot}$ and below as indicated by \cite{LeBrun_2014MNRAS,McCarthy2010MNRAS} etc. Cluster core properties can also be effected by AGN feedback by quenching star formation, thus resulting change in X-ray emissions and related scaling laws \citep{Rasia_2014ApJ,LeBrun_2014MNRAS}, therefore, becomes a subject of our future research. Above studies also show that both coolSF and coolSF+AGN feedback are producing far better results than the non-radiative simulations. Self similar parameters produced by coolSF and additional AGN feedback are by and large agree with the observations. With the data set they have compared, it is very difficult to conclude which one is fitting better. For some parameters though, either of the model is slightly over or under-predicting, and few specific parameters only are better produced by coolSF+AGN simulations.

In current study, coolSF model of our's itself is seen to be fairly able to reproduce the observations (detail study will be followed in the next sections), unlike the studies by \cite{LeBrun_2014MNRAS,McCarthy2010MNRAS}. We have thus used the coolSF model for creating our sample set of about 360 objects and an extensive work has been done to figure out a possible break away point or knee in cluster scaling laws those distinguish correctly a `group' from a cluster. The reasons behind the unique properties of groups have also been investigated. 

Section~\ref{intro} deals with the introduction to the problem with its background motivations. The simulation details and sample selection etc. has been written in Section~\ref{sample}. Details of studied parameters and their results are given in Section~\ref{study-self}. For reliability of our simulations, we have presented a resolution study in Section~\ref{res-study}. Discussion on the distinct features and point of segregation of groups from the clusters has been written in Section~\ref{discus}. Finally, we have summarised our findings in Section~\ref{conclude}.

\section{Simulation details and sample selection}\label{sample}

Lack of studies on properties of low mass systems prompted us to model them using cosmological simulations. To create our sample set, basic simulations were performed with the adaptive mesh refinement (AMR), grid-based hybrid (N-body plus hydro-dynamic) code Enzo v.~2.2 \citep{Bryan_2014ApJS}. With introduction of 2 nested child grid and another 4 levels of AMR at the central (32 $\rm{Mpc})^3$ volume (i.e. total 6 levels of refinement), we have achieved a resolution of $\sim$ 30 kpc at the highest level for our studied sample set. For resolution study, we have simulated two other sets of data with a lower resolution  (`LOWRES' here after) $\sim$ 60 kpc and a higher resolution (`HIGHRES' here after) $\sim$ 15 kpc than the reference set (`REFRES' here after) i.e. $\sim$ 30 kpc. As cosmological parameters, we have taken a flat $\Lambda$CDM background cosmology with $\Omega_\Lambda$ = 0.7257, $\Omega_m$ now = 0.2743, $\Omega_b$ = 0.0458,  h = 0.702 and primordial power spectrum normalization $\sigma_8$ = 0.812 \citep{Komatsu_2009_APJS}. 

In an AMR simulation, it is very much important to resolve the objects very well. Density has been used as the primary refinement criteria to resolve the cells in this study. If a cell has density 4 times the average density of the neighbouring cells, the cell is refined. Similarly, if a cell has 4 times higher dark matter mass than the average mass of its neighbouring cells, the cell is refined.  Since, density is very low at the outskirts of the large scale objects, we also require a dynamical parameter that covers almost all the volume of the objects. It is therefore convenient to choose the shocks as second parameter to refine the cells as, in forming objects, shocks spans over most of the volume. This is also justified since, shock heating is a primary heating engine in large structures that helps objects resist the rapid collapse due to radiative cooling and decrease unphysical star formation activity. Proper shock capturing is thus very important for scaling studies. In this study, shock computing has been done using un-split velocity jump method of \citet{Skillman_2008ApJ} with a temperature floor of $\rm{T}^4$K which is found to give better results in AMR simulations \citep{Vazza_2011MNRAS}. Heating, cooling and feedback physics has a dominant effect on smaller structures. We have implemented radiative cooling due to X-ray, UV \& optical emissions and heating is due to stellar motions and Supernova \citep{Sarazin1987ApJ}. Star formation and feedback scheme of \citet{Cen1992ApJ} have been implemented with a feedback of 0.25 solar. This is the model of additional physics that we have named as `coolSF' runs. 

After generating initial conditions using the Eisenstein \& Hut transfer function \citep{Eisenstein_1998ApJ}, hydro plus N-body simulations were performed from redshift z=60 to current redshift z=0. We have saved data outputs in the intermediate redshifts and a very frequent snapshots were taken below redshift 1. Physical parameters are computed on these snapshots using tools developed from yt \citep{Turk_2011ApJS}.

We have simulated 10 realizations of parts of our universe, each of volume $(128\;\rm{Mpc})^3$. Central $(32\;\rm{Mpc})^3$ of highly resolved volume of each simulation contains few 10s of groups and clusters. We have identified individual objects using yt halo finder tool. For identifying the objects, we have used virial radius to be their delimiter, where, virial refers to the quantity at over density of 200 to the critical density of the universe at that redshift. In the further text, over density should be understood as compared to the critical density only. Finally, about 360 objects have been selected in the mass range of $5\times10^{12}\; M_{\odot}$ to 2.5 $\times 10^{15}\; M_{\odot}$. Our mass resolution at the smallest child grid is $<$ $10^{9}\; M_{\odot}$ providing enough mass resolution even for the groups with mass $5 \times 10^{12}\; M_{\odot}$ . Also, with $\sim$ 30 kpc spatial resolution, systems above this mass, having virial radii above 500 kpc get adequately resolved in space. 

\section{Study of self similarity in the selected sample set}\label{study-self}

The scaling laws of self similarity are derived from ideal spherical collapse model with gravity only situations. Dark Matter (DM) only studies that corroborate the results of existence of self similarity \citep{kaiser1986MNRAS,Navarro_1996ApJ,Vikhlinin2006ApJ} could not explain the observed $\rm{L_x}$-T scaling for smaller objects. This model has failed to account for the dynamical and transient events such as mergers, also for thermodynamics and other forms of energies in these systems. So, the introduction of baryons to the DM-only formalism changes the physical conditions significantly, causing substantial variations in the energetics of these objects \citep{Dai_2010ApJ,Bharadwaj2015}. Since, the overall environment and dynamical processes experienced by the smaller and larger systems are different (see Section~\ref{intro}), it is quite possible that smaller objects can deviate from the cluster scaling laws. So, to understand the energetics of groups and clusters, we certainly need to study the correlation scaling of different physical parameters, both thermal and non-thermal along with the hydrostatic ones. Further, transformation of thermal to non-thermal energy could be a strong parameter to check for, as this indicates the ongoing dynamical processes that control the energy budget of the systems, thus explaining the deviation from self-similarity. 

\subsection{Self similarity in thermal properties}\label{self-sim}

Dark matter being the dominant matter component of the universe, gas would just follow the DM collapse. In such a situation, as derived by \cite{Peebles_1980lssu.book,Kitayama_1996ApJ,kaiser1986MNRAS}, relations among the observables of gas properties with the mass observables can in fact be formulated. Virial temperature (T) of the system can be related to the total virial mass, by considering the heating of the medium through conversion of potential energy of collapsing gas as $\rm{T\propto (\Delta \rho_{cr})^{1/3} M^{2/3}}$. Assuming thermal Bremsstrahlung X-ray emission to be $\propto\rho \rm{T^{1/2}}$, X-ray luminosity will scale to the virial temperature as $\rm{L_X \propto\rho^{2}  T^{1/2} r^{3}}$ i.e. $\propto \rm{T^{2}}$ as $\rm{r^3 \propto M \propto T^{3/2}}$. These also indicate that two very closely related parameters i.e. mass and X-ray luminosity should have a very tight correlation of $\rm{L_X \propto M^{4/3}}$, when estimated from independent observations. Entropy  (S) being a function of density and temperature, should also show self similar relation which is given by $\rm {S \propto T}$. So, the crucial and independent self similar scaling relations needs to be studied are $\rm{L_X}$-M, $\rm{L_X}$-T and S-T and parameters must be computed within virial radius, i.e. within over-density of $\sim$ 200.

In our simulations, mass and virial temperature has been computed upto the over-density radius of $\rm{r_{200}}$ i.e. usual virial radius of the systems. We have computed virial temperature as a function of mass and virial radius of the object i.e. $\rm{^1T_{vir} = \frac{G M \mu m_{p}}{(k_b R_{vir})}}$ (operational temperature definition 1). For a bound and self-gravitating system, the virial mass can be related to the velocity dispersion of the system by $\rm{M = \frac{3 R_{vir} \sigma^{2}}{G}}$. Since, velocity dispersion $\sigma$ can be computed independently in our simulations by filtering out the bulk motion, we get a second definition of temperature as $\rm{^2T_{vir} = \frac{3\mu m_{p} \sigma^2}{k_b}}$, which is the same as the X-ray temperature if $\beta_{spec}$ value is considered to be unity (observed average value \citep{Girardi_1996ApJ}) in $\rm{T_{x} = \frac{3\mu m_{p} \sigma^2}{k_b \beta_{spec}}} $. Finally, entropy (S), which is a function of temperature and density of the objects, has been computed using the relation $\rm{ \frac{k_b T_{vir}}{\mu m_p \rho^{\gamma}}}$. 

Gravitational collapse, mergers during structure formation leads to adiabatic compression of the intra-cluster medium (ICM) and heating up the structures to as high as $10^8 \rm{K}$ \citep{Sarazin_2002ASSL,Mathis_2005MNRAS,Paul2011ApJ}. Such a medium then emit X-rays through thermal bremsstrahlung (For a review:\cite{Felten_1966ApJ}), Inverse Compton Scattering (ICS) of Cosmic Microwave Background Radiation etc. \citep{Costain_1972ApJ}. Values of X-ray luminosity, emissivity, and photo emissivity fields for a given photon energy range can be obtained using Cloudy~\citep{Ferland_1998PASP} code. For the present study, we have limited the energy of X-ray emission between 0.1 keV to 12.0 keV in order to comply with the range of existing X-ray telescopes. This range is also sufficient for calculating approximate bolometric luminosity as beyond this range X-ray flux from thermal gas in galaxy clusters falls by $10^{-3}$ \citep{Henriksen_1986ApJ} times i.e. the considered energy band covers almost 99\% of the bolometric flux of the cluster samples. 

\subsubsection{Comparison of simulated parameters with observations}

\begin{figure*}
\includegraphics[width=9cm]{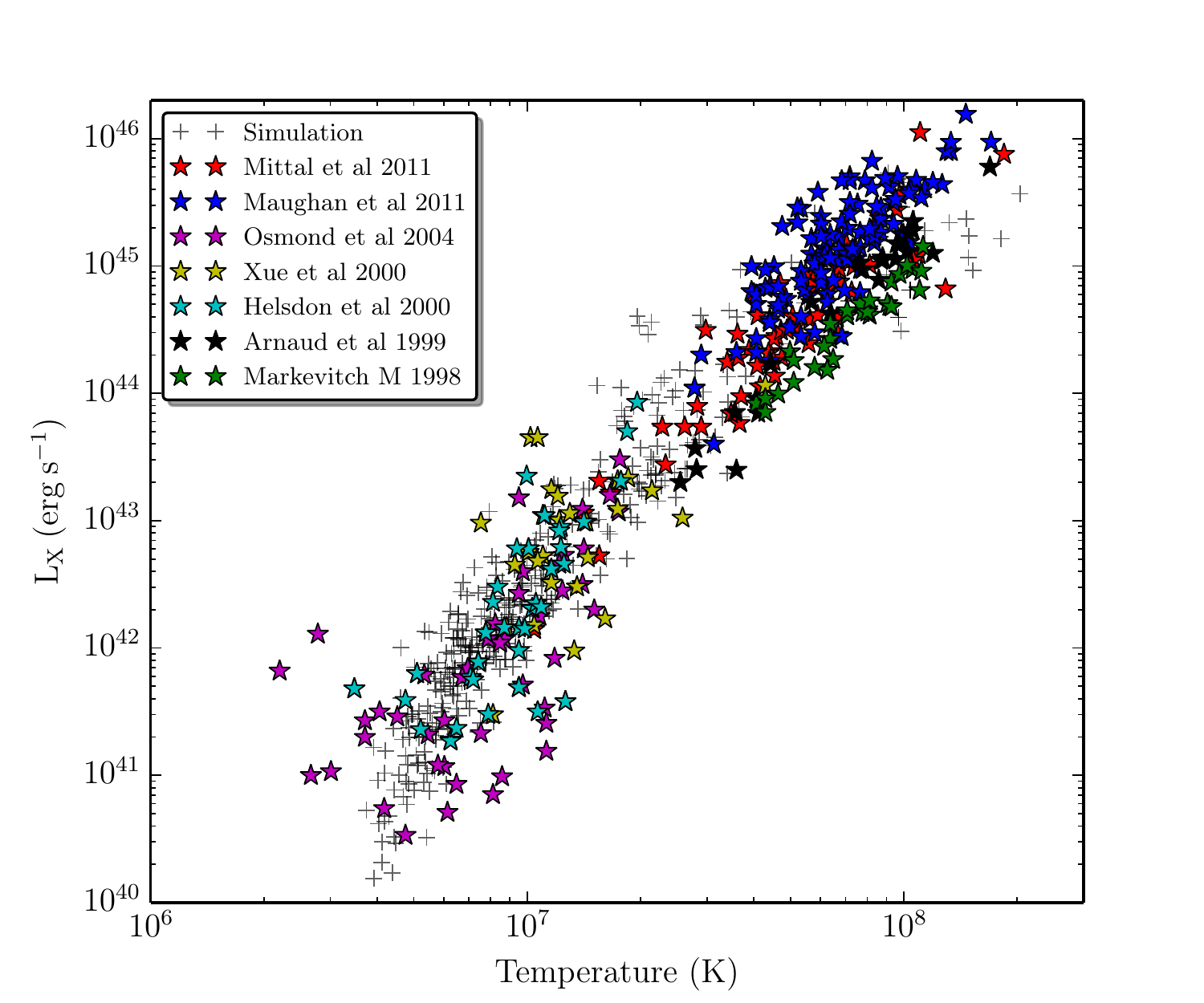}\hspace{-0.8cm}
\includegraphics[width=9cm]{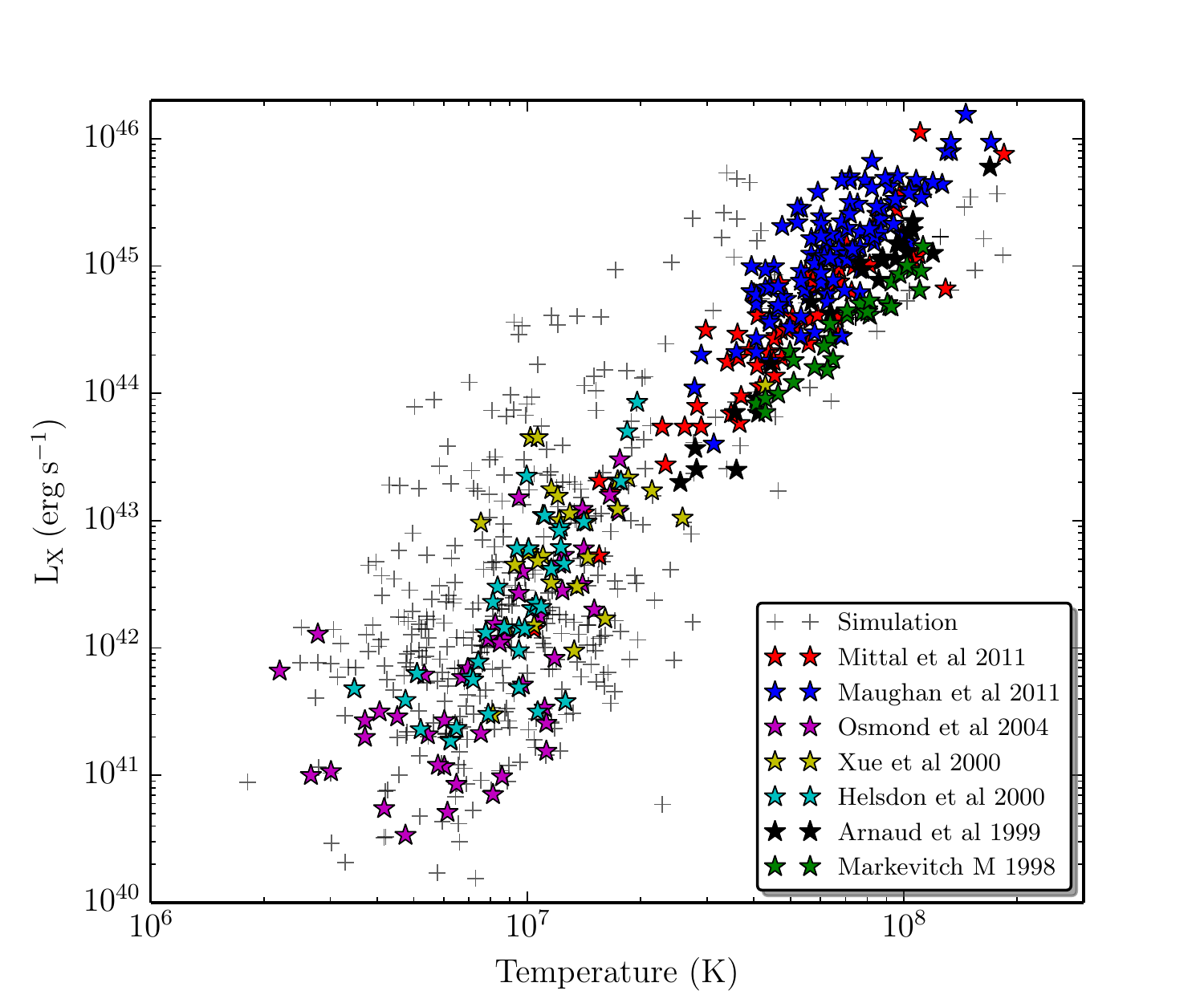}
\caption{{\bf Panel 1:} X-ray luminosity plotted against virial temperature (definition 1, $\rm{^1T}$) computed upto radius $\rm{r_{500}}$. Observed data points from \citep{Mittal_2011A&A,Helsdon_2000MNRAS,Maughan_2012_MNRAS, Markevitch_1998ApJ,Osmond_2004_MNRAS,Xue_2000ApJ,Arnaud_1999MNRAS,Zhang_2011A&A} have been overplotted as coloured stars. {\bf Panel 2:} Same plot as above but, against virial Temperature (definition 2, $\rm{^2T}$).}\label{x-ray-obs}
\end{figure*}

Most of the X-ray observations are done upto over-density ratio of 500 (i.e. till $\rm{r_{500}}$). So, to validate our results using observations, we have created a data set by computing X-ray luminosity  till $\rm{r_{500}}$.  X-ray luminosity has been plotted against virial temperature with both $\rm{^1T_{vir}}$ and $\rm{^2T_{vir}}$ of the selected samples from our simulations in Figure~\ref{x-ray-obs}, panel 1\&2 respectively. Observed X-ray luminosity from similar objects found in different studies have been over-plotted in the same. Since, both virial temperatures have been derived from hydrostatic equations that are related to self similarity, we have chosen to plot X-ray luminosity against these temperatures. This is also relevant as almost all the observed objects that are over-plotted here are available against virial temperature only. It can be noticed that simulated mock data set nicely follow the observed data trend.  A large scatter in $\rm{L_x}$-T plot can be noticed when plotted against $\rm{^2T_{vir}}$, which is a function of velocity dispersion. Since, we did not put any bias while selecting our sample set from our simulations, it is obvious that it will contain merging objects as well. Velocity dispersion being a strong function of dynamics of the objects, mergers can strongly influence the temperature calculated from this parameter. On the other hand, possibly mass being comparatively smoothly varying parameter, shows less scatter. So, to have a better control over computation of scaling relations, we have chosen $\rm{^1T_{vir}}$ for further temperature related calculations.

\begin{figure*}
\includegraphics[width=9cm]{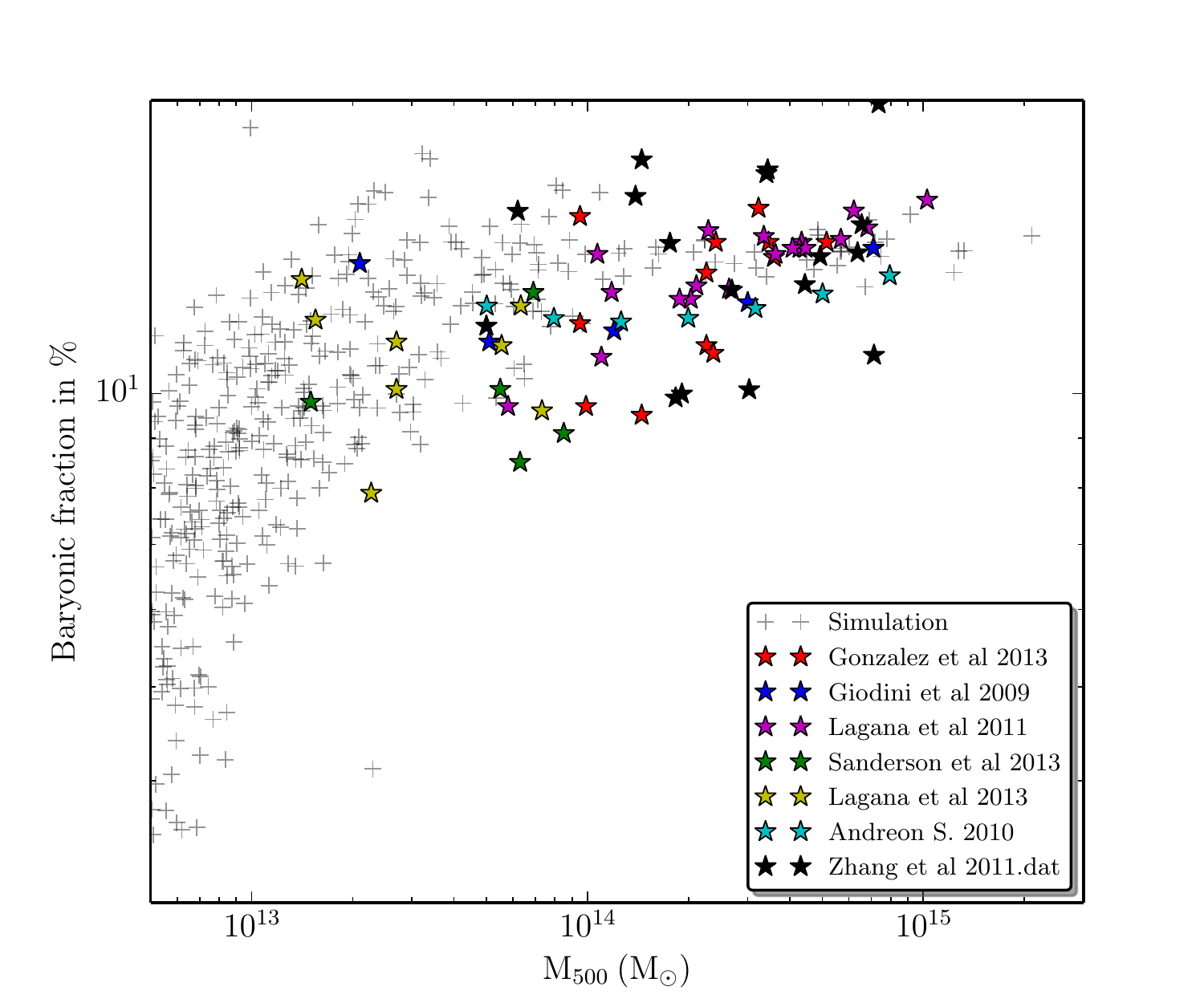}\hspace{-0.8cm}
\includegraphics[width=9cm]{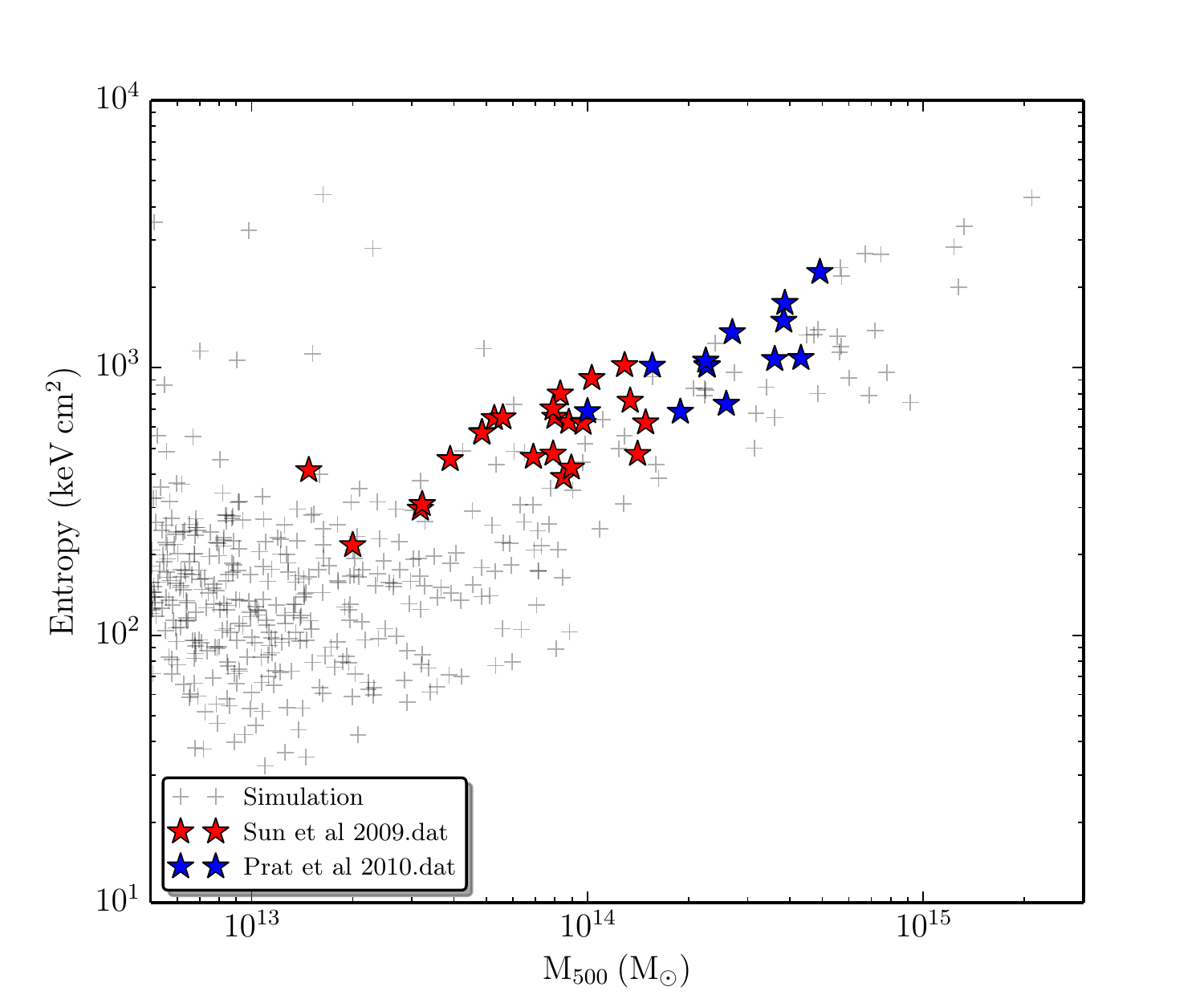}
\caption{{\bf Panel 1:} Baryon fraction (in \%) has been plotted against mass ($\rm{M_{500}}$). Observed data points from \citet{Gonzalez_2013ApJ,Giodini_2009ApJ,Lagana_2011ApJ,Sanderson_2013MNRAS,Lagana_2013A&A,Andreon_2010MNRAS} have been overplotted as coloured stars. {\bf Panel 2:} Entropy against virial mass has been plotted \citep{Sun_2009ApJ,Pratt_2010A&A} and compared with the overplotted observed data.}\label{x-ray-obs1}
\end{figure*}

Further, we have plotted baryon fraction and entropy within $\rm{r}_{500}$ against mass ($\rm{M}_{500}$) for our simulated samples and over plotted the available set of observed data in Figure~\ref{x-ray-obs1}, panel 1\&2 respectively. Though, simulated baryon fraction seems to follow observed values well, entropy shows a deviation below $\sim 10^{14} \; M_{\odot}$. Lack of observed data points in low mass makes it difficult to understand the reason for the  apparent deviation. So, overall, these three parameters fits well to the observations and validate our coolSF simulation model used in this study to a large extent. It can also be noticed that, both observed and simulated data set show strong indication of deviation from cluster scaling for low mass systems. So, further, in this paper, we will compute the point of break where cluster scaling deviates from the self similarity.

\subsubsection{Breaks in self similar laws}

\begin{figure*}
\includegraphics[width=6cm,trim={0.1cm 0.1cm 0.2cm 0.3cm},clip]{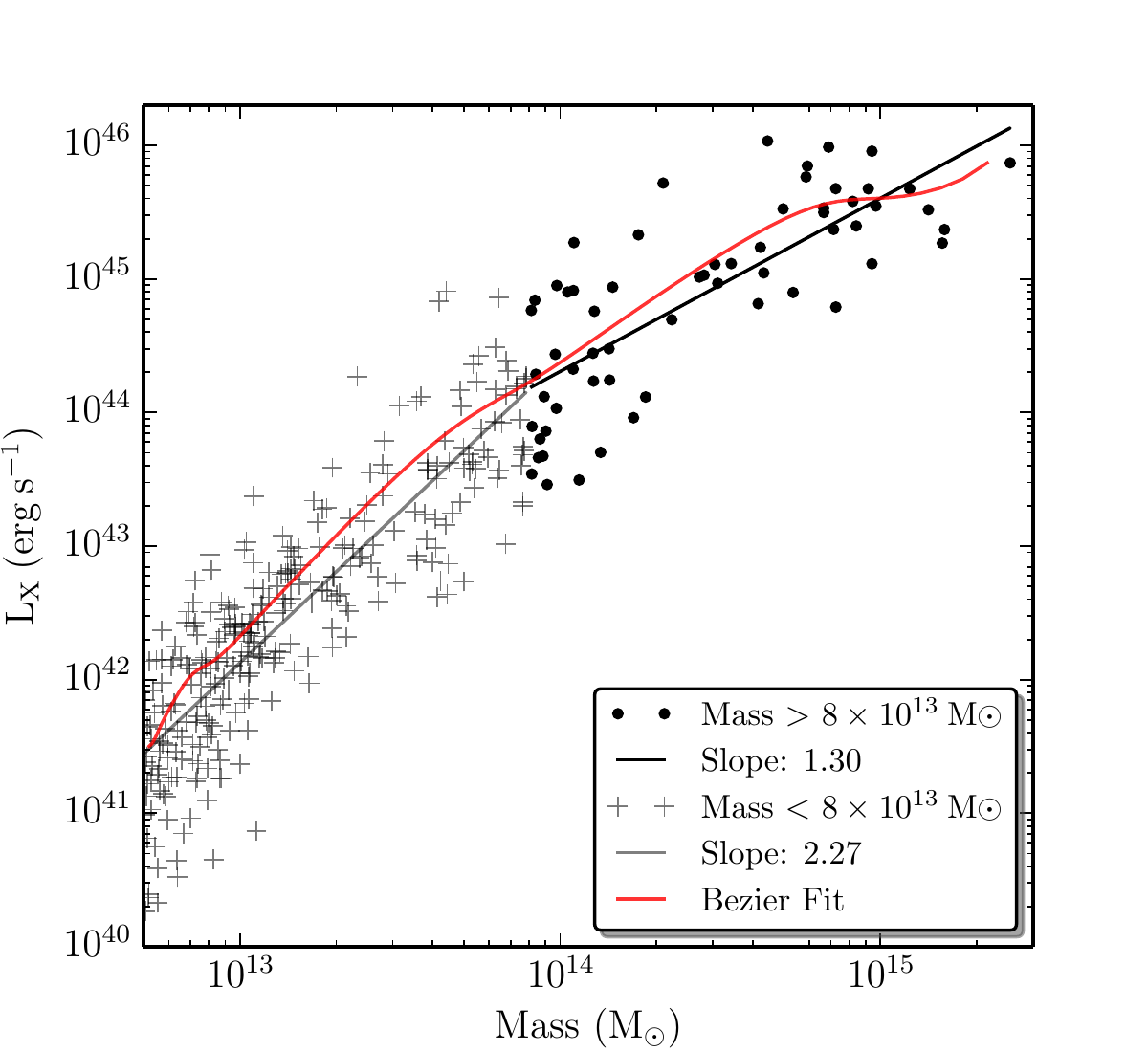}\hspace{-0.4cm}
\includegraphics[width=6cm,trim={0.1cm 0.1cm 0.2cm 0.3cm},clip]{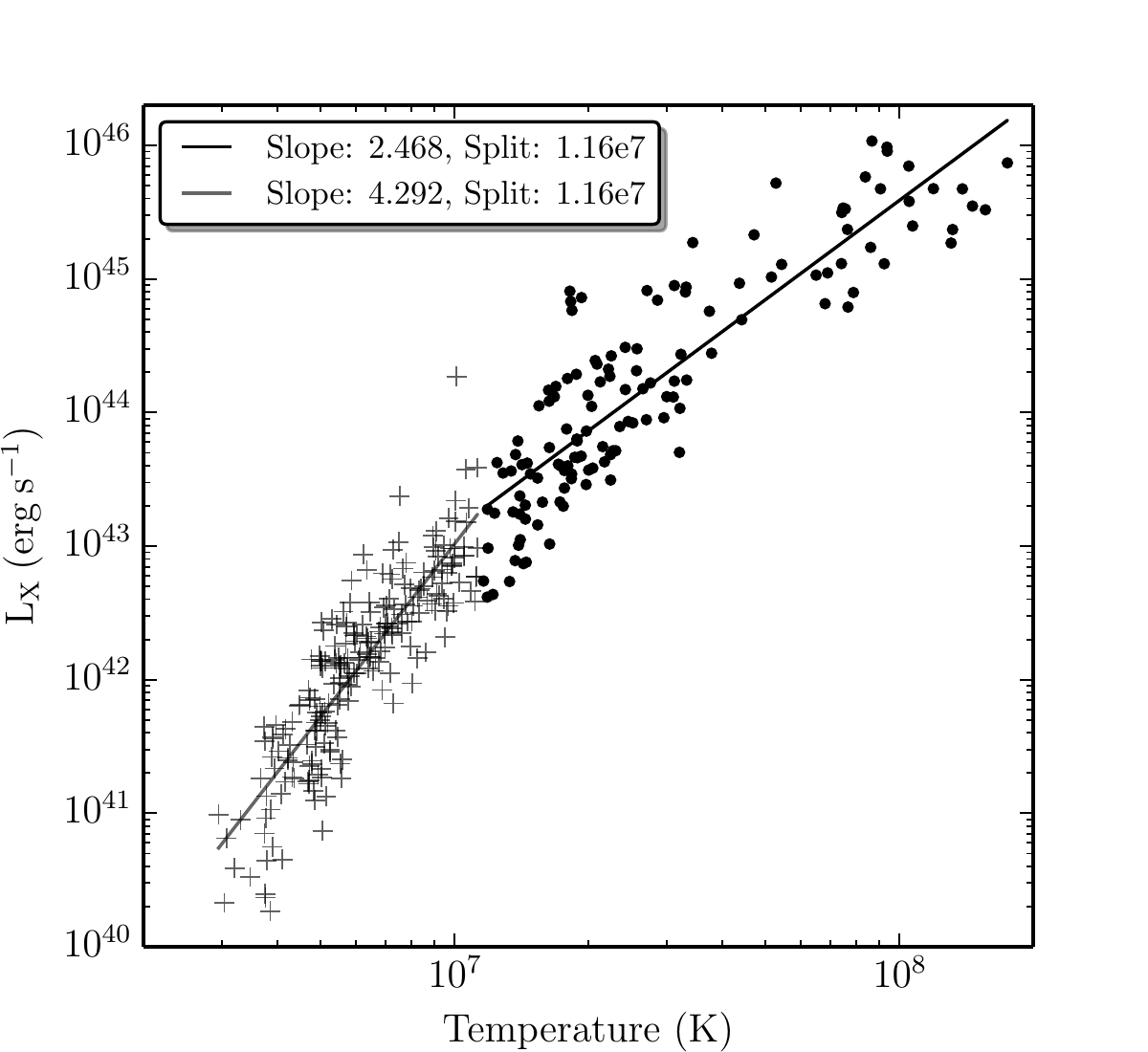}\hspace{-0.4cm}
\includegraphics[width=6cm,trim={0.1cm 0.1cm 0.2cm 0.3cm},clip]{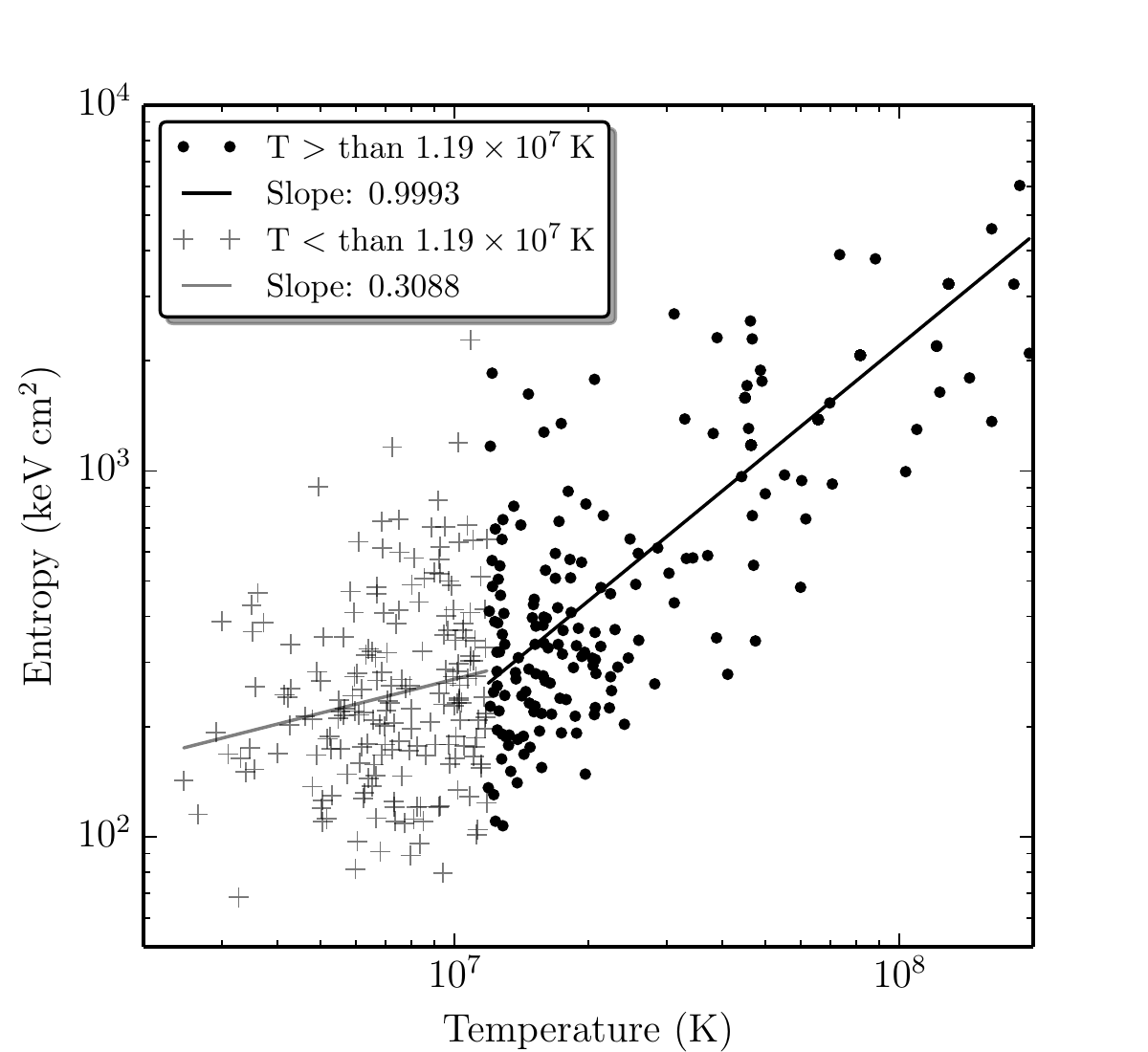}\hspace{-0.4cm}
\caption{X-ray Luminosity vs virial mass and virial temperature (upto $\rm{r_{200}}$) has been plotted from our sample data set in Panel 1 \& 2 respectively. B\'ezier curve (Section~\ref{self-sim}) have been fitted and plotted over data in Panel 1. In Panel 3. entropy of these systems plotted against virial temperature. All the plots are fitted with two slopes with break points using method as specified in Section~\ref{self-sim}.}
\label{xray-mass}
\end{figure*}

Self similar scaling laws are derived for the hydrostatic equilibrium or virial equilibrium. From the spherical collapse models, the original formulation was done using virial radius at over-density of $\sim 200$ (more accurately, 178) \citep{kaiser1986MNRAS,Cole_1996MNRAS,Miniati_2015Natur}. Though, use of over-density of $\sim 200$ would be more appropriate, most of the previous studies are observed to be done with over density 500 or $\rm{r}_{500}$ which does not guaranty a hydrostatic  equilibrium. So, to ensure correct formulation and appropriate calculations, we have computed our parameters at over-density 200. 

Figure~\ref{xray-mass}, Panel~1 shows the X-ray luminosity against mass within the radius $\rm{r_{200}}$ obtained from our simulated data set. It is always difficult to understand actual trend from the data set when large fluctuations are present and is a multivalued function. To overcome this difficulty, we have taken the statistical average of the data at each point and fitted a cubic B\'ezier curve to smooth out the fluctuations to compute actual trend and the bending points in the curve. In Panel 1, the B\'ezier fitted curve seen to turn towards cluster self similarity at $\sim$ $5\times 10^{13}\;\rm{M_{\odot}}$ and falls into the cluster scale beyond $\sim 8\times 10^{13}\;\rm{M_{\odot}}$. Further, to estimate the accurate point of break in the scale, we have computed the fitting power laws using all the data points in such a manner that slopes get connected at the break point. Power law fitting in figure~\ref{xray-mass}, Panel~1 shows that structures with mass below $\sim 8\times 10^{13}\;\rm{M_{\odot}}$ follows a 9/4 scaling, whereas structures above this mass has a scaling of 4/3 i.e. $\rm{L_X \propto M^{4/3}}$ as expected from the self similarity of clusters discussed in the Section~\ref{self-sim}. 

X-ray luminosity with virial temperature indicates a break at around $1.16\times10^7$ K i.e. 1 keV in B\'ezier curve. Power law when fitted to the hotter to cooler systems in Figure~\ref{xray-mass}, Panel~2, a break point at 1 keV has been observed. The hotter objects are found to follow a scaling of $\rm{L_x \propto T^3}$ exactly as observed in \cite{Allen_1998MNRAS}, whereas cooler objects show a scaling of  $\rm{T^{3/2}}$. Further, entropy of the selected samples are plotted against virial temperature in Figure~\ref{xray-mass}, Panel 3. Slope of S-T curve has been found to be $\sim$1 for objects with temperature above 1 keV. But, if we consider objects below 1 keV, it becomes much flatter with a slope of 0.3. Fluctuations in the data below this temperature also indicate that they have non correlated entropy distribution. Strikingly, this also shows a power law type behaviour only above the temperature of 1 keV as in the case of X-ray luminosity.

\subsection{Baryon fraction evolution}\label{bar-frac}

\begin{figure*}
\includegraphics[width=6cm,trim={0.1cm 0.1cm 0.2cm 0.2cm},clip]{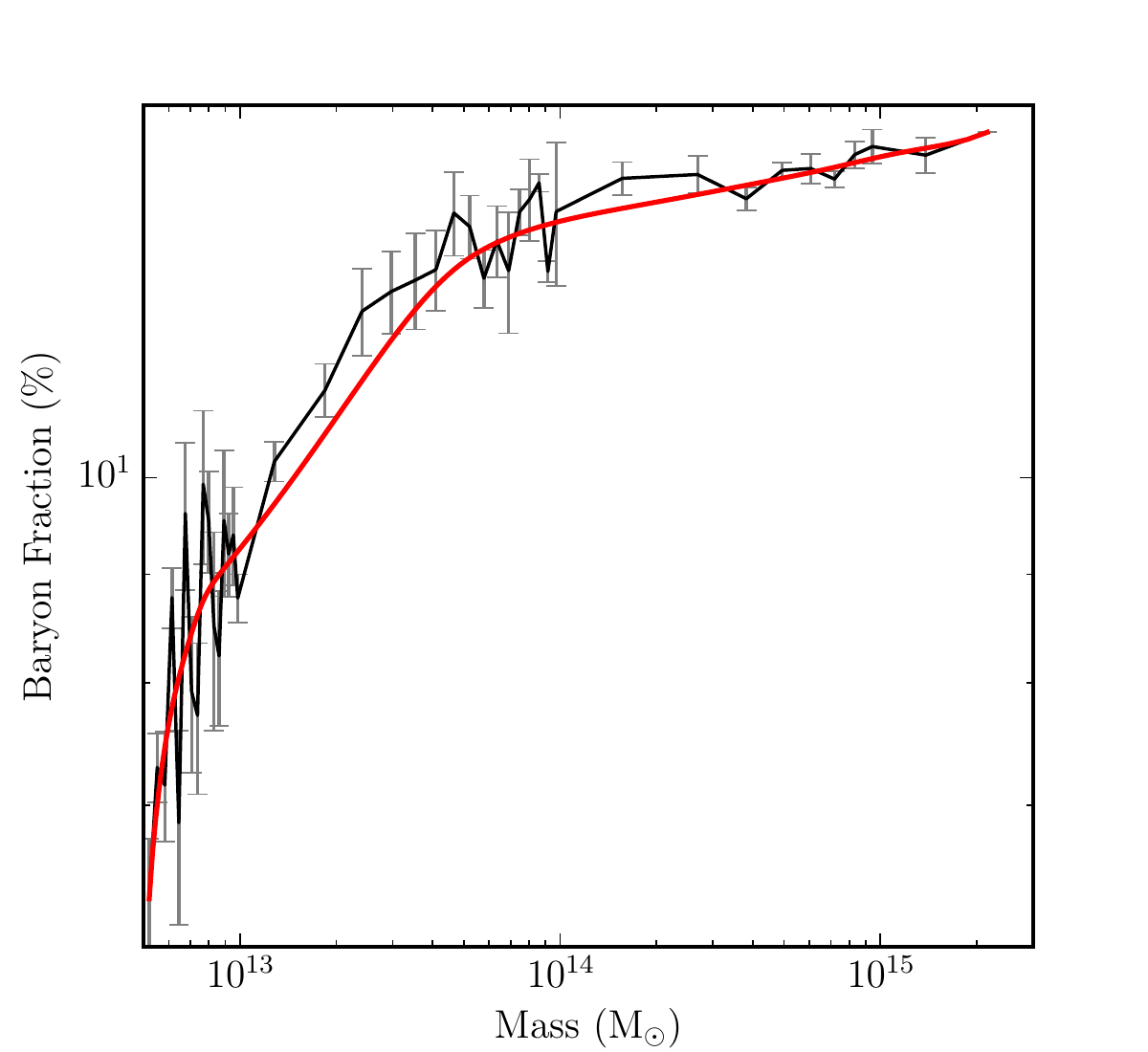}\hspace{-0.4cm}
\includegraphics[width=6cm,trim={0.1cm 0.1cm 0.2cm 0.2cm},clip]{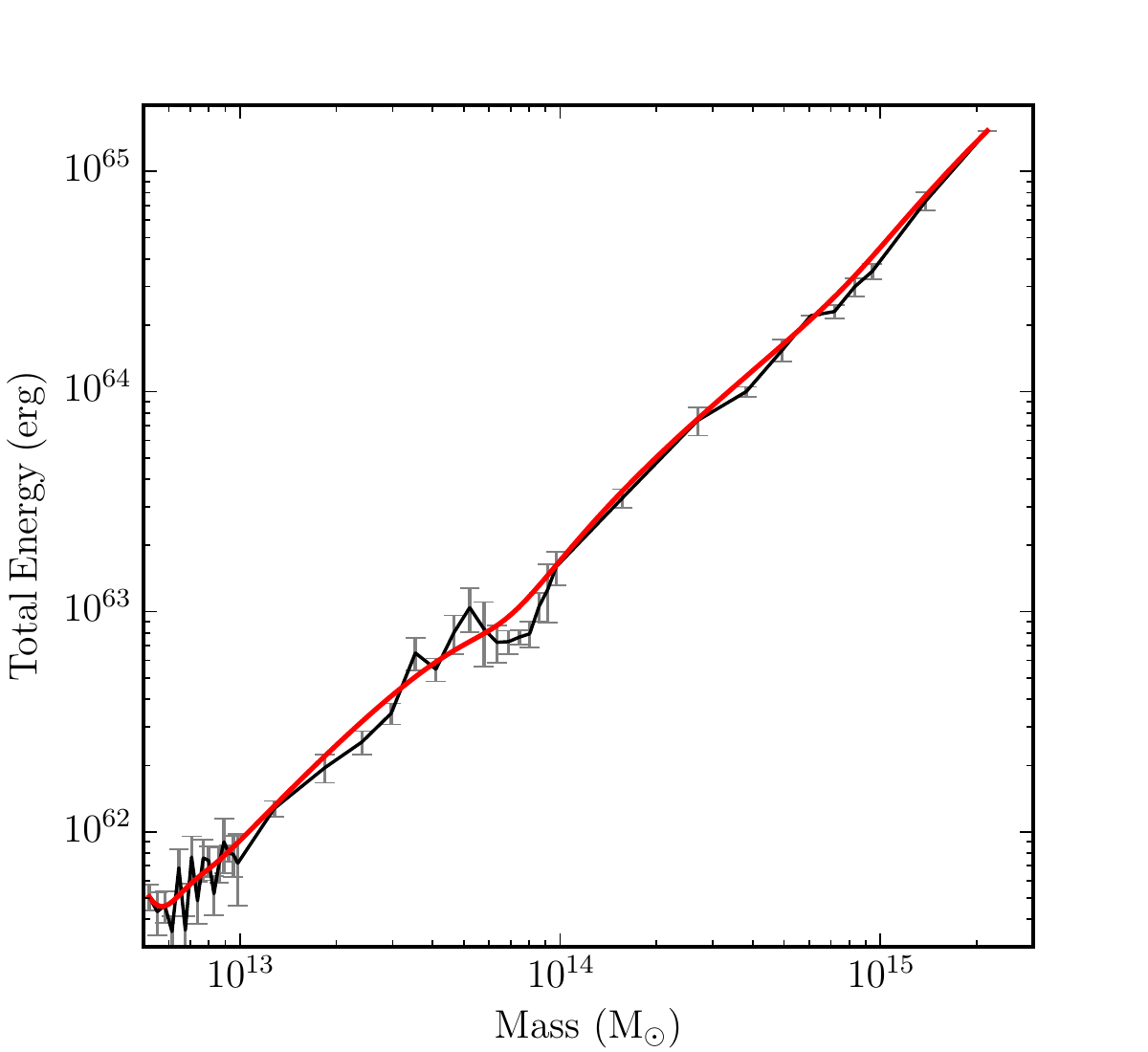}\hspace{-0.4cm}
\includegraphics[width=6cm,trim={0.1cm 0.1cm 0.2cm 0.2cm},clip]{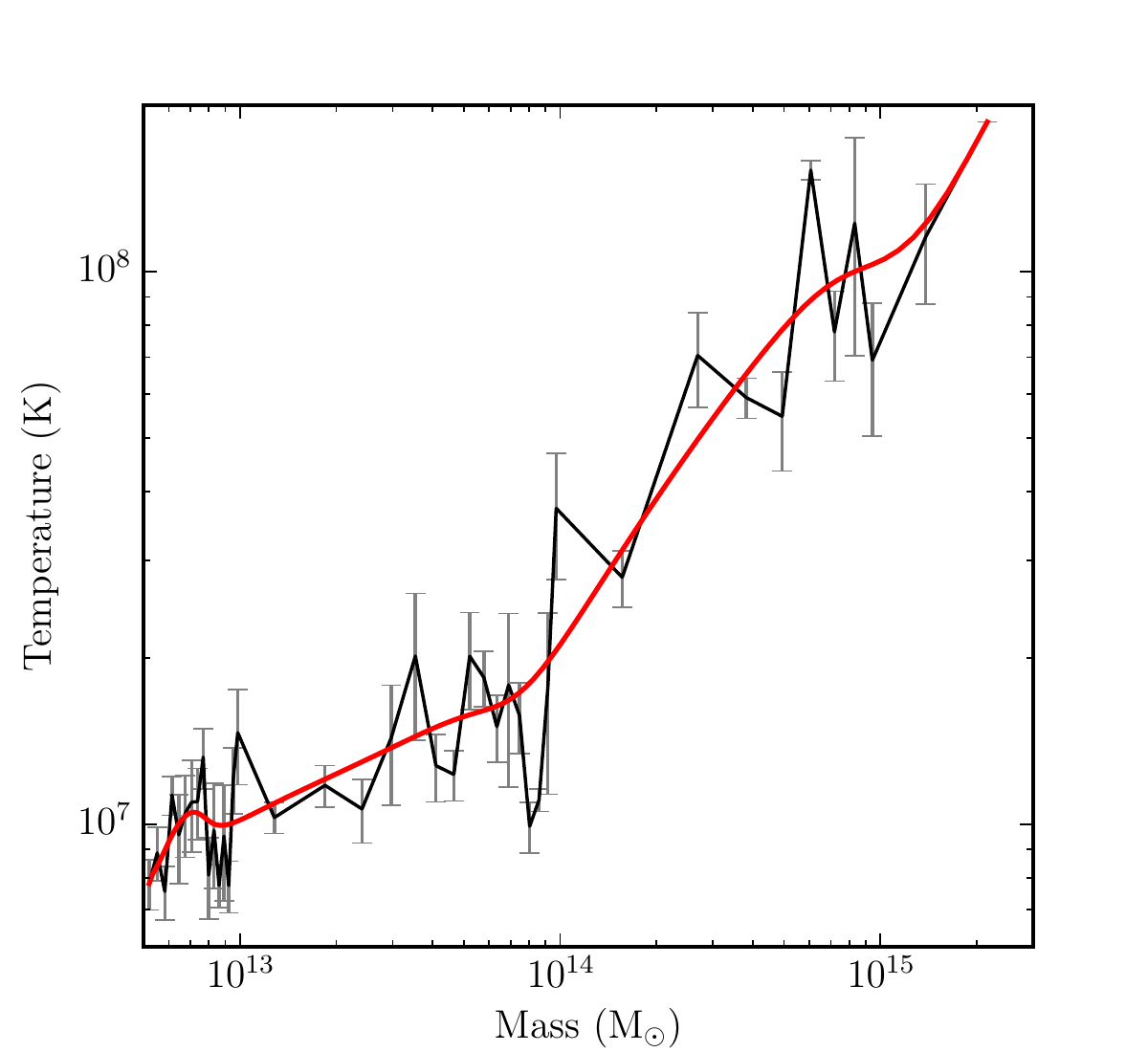}\\
\caption{Averaged baryon fraction, total energy and virial temperature computed upto $\rm{r_{200}}$ have been plotted against virial mass in the Panel 1, 2 \& 3 respectively and fitted with B\'ezier curve as mentioned in Section~\ref{self-sim}.}
\label{baryon-frac}
\end{figure*}

Thermally interacting gas (baryons) that has very less fractional abundance in the universe, brings a huge change in observable energy budget of these massive structures. Hot baryons are a very crucial component that controls the X-ray emission as well as all kinds of non-thermal emissions. Thus, understanding the evolution of gas in these systems is very crucial for describing the hierarchical growth of structures in the universe. We have computed the fractions of baryons within the virial radius ($\rm{r_{200}}$) of the chosen structures spanning almost 3 order of mass (see Figure~\ref{baryon-frac}), Panel 1. The fitted line in Figure~\ref{baryon-frac} has two clear breaks, indicating a drastic change of properties in the systems across these points. The first knee of the curve is at $\sim 6-8\times 10^{12} \; \rm{M_{\odot}}$, while the other one is at around $6-8\times 10^{13}\; \rm{M_{\odot}}$. It has been observed that the baryon fraction in structures below mass $8\times 10^{13}\; \rm{M_{\odot}}$ has a very steep decrement and oscillatory in nature indicating a rapid change in its properties. Above this mass, baryon fraction almost gets stabilized at $\sim$ 14\% and a very flat slope has been observed.  

\subsection{Total Kinetic energy distribution}\label{tot-en}

Deciphering the kinetic energy in LSS needs a proper understanding of energy shared among baryons and DM. During structure formation, a large amount of binding energy converts to kinetic energy. Released binding energy is shared by the DM and baryons and changes the ICM dynamics significantly. Total kinetic energy in such a system is given by KE = $\rm{K_g + K_d}$ =$\frac{3}{2}\;\sigma^{2}_v \left(\rm{ M_g  + M_d}\right) \;$, \citep{Diaz2010MNRAS}. Where, $\rm{M_g}$ and $\rm{M_d}$ are the gas and the dark matter mass respectively, while, ${\sigma}_v$ is the radial velocity dispersion assumed to be the same for the DM and the galaxies i.e baryons. But, as it has been observed in our study (Section~\ref{bar-frac}), baryon fraction across the knees varies largely and their thermal interaction should also be different, thus, approximation of same velocity dispersion for DM and baryon could lead to an erroneous result. We have thus computed baryon and DM velocity separately (i.e. $\sigma^{2}_b$ and $\sigma^{2}_{dm}$) and modified the above equation accordingly as $\rm{E}_{tot} = \left( \rm{M_g \sigma^{2}_b + M_d \sigma^{2}_{dm}} \right)$. Total energy of the system has been plotted in Figure~\ref{baryon-frac}, Panel 2, also indicate a similar dichotomy like in the other parameters studied in this paper. Strikingly, again the turning point is found to be at $\sim 8\times 10^{13}\; \rm{M_{\odot}}$.

\subsection{Self similarity in non-thermal properties}

\begin{figure*}
 \includegraphics[width=6cm,trim={0.1cm 0.1cm 0.2cm 0.2cm},clip]{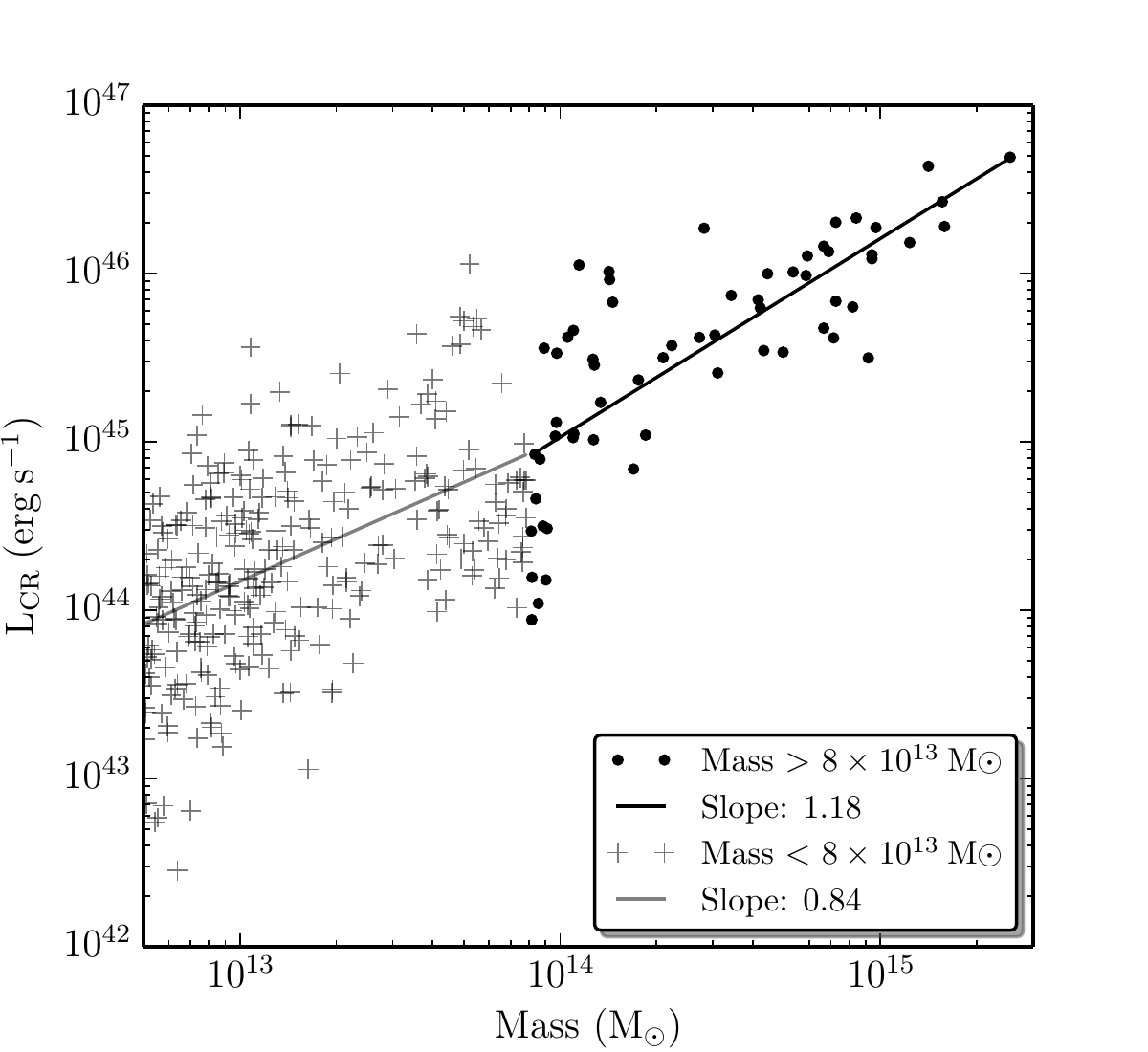}\hspace{-0.4cm}
\includegraphics[width=6cm,trim={0.1cm 0.1cm 0.2cm 0.2cm},clip]{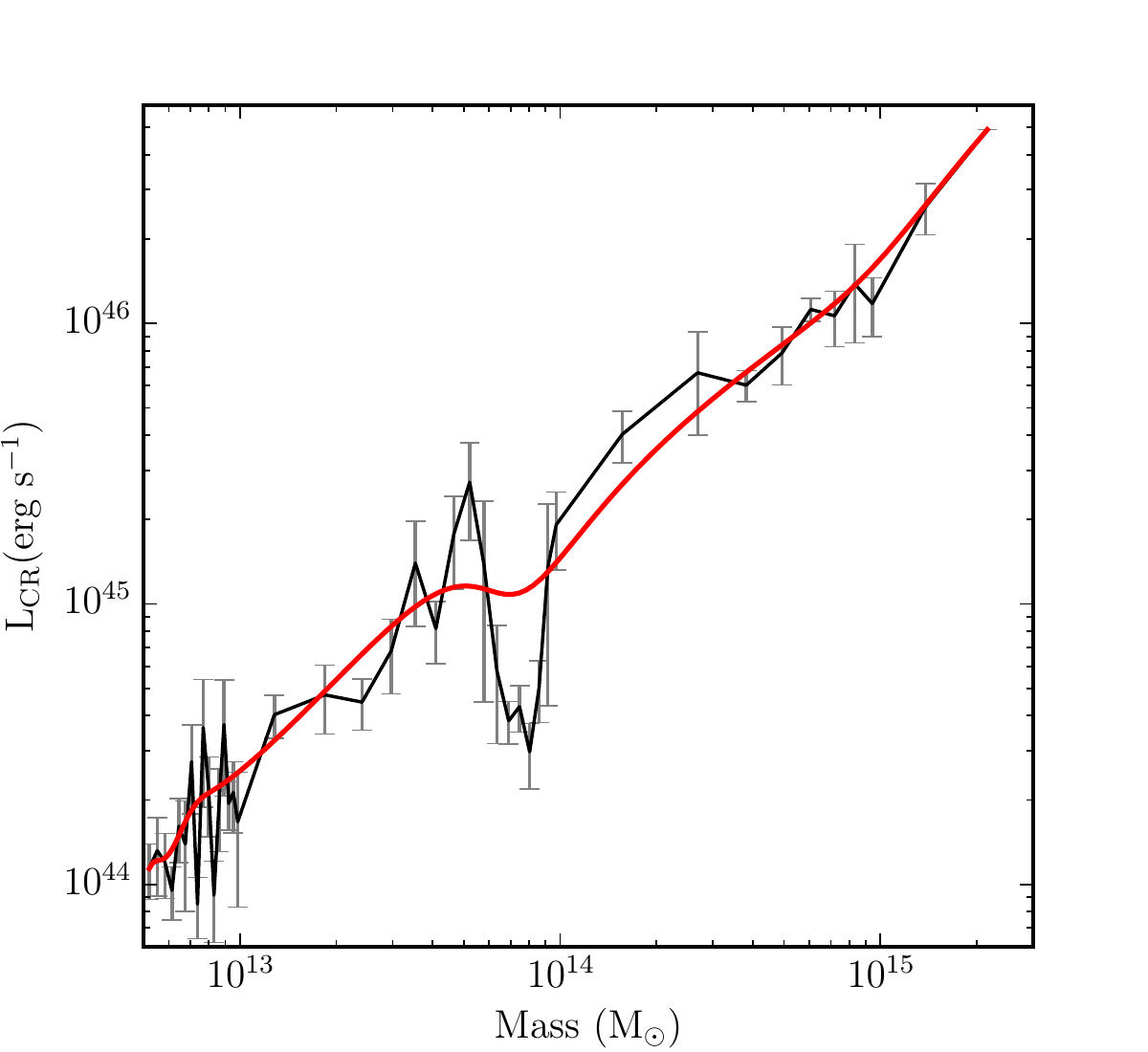}\hspace{-0.4cm}
\includegraphics[width=6cm,trim={0.1cm 0.1cm 0.2cm 0.2cm},clip]{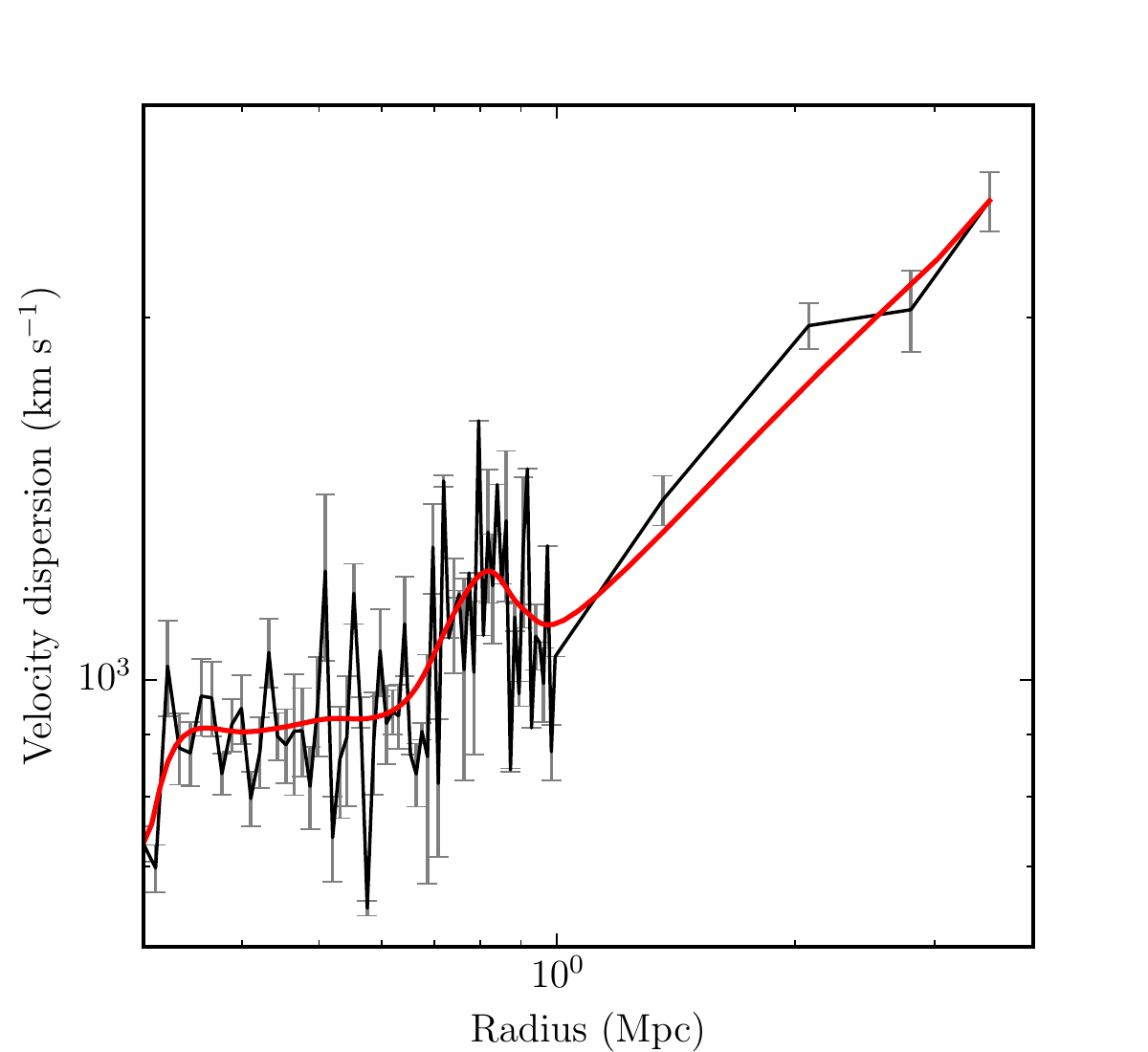}\hspace{-0.4cm}
\caption{ In Panel 1, cosmic ray luminosity ($\rm{L_{CR}}$) has been plotted against virial mass and slopes are computed similar to Figure~\ref{xray-mass}. Similarly, B\'ezier curves are fitted to ($\rm{L_{CR}}$) vs mass and Total velocity dispersion (DM plus baryon) vs virial radius ($\rm{r_{200}}$) in Panel 2 \& 3 respectively.}
\label{CRs-scale}
\end{figure*}

Though, intra cluster medium (ICM) is dominated by thermal particles a non-negligible fraction of non-thermal cosmic-ray (CR) particles has also been observed. Shared ICM energy attains an equipartition at a scale larger than dissipation scale, if ICM gets enough relaxation time (i.e. $\epsilon_{therm} \sim \epsilon_{CR} \sim E_B \sim \epsilon_{turb}$ \citep{Longair1994}. The most effective mechanism active for particle acceleration in LSS is the diffusive shock acceleration (DSA, \cite{Drury1983RPPh}). The pool of energetic particles that exists in ICM due to AGN activity, star formation, supernova explosions etc. are injected at the shocks along with the thermally energised particles. Particles get accelerated by DSA to cosmic rays (CRs). CR flux is a strong function of Mach number and can be computed from the hydrodynamic parameters using the function obtained by \citet{Kang_2013ApJ} as $f_{CR}= \eta(\mathcal{M}) \times \frac{1}{2}\rho(\mathcal{M} c_{s})^3$. Where, cosmic ray acceleration fraction is a function of Mach number i.e. $\eta(\mathcal{M})$. From Figure~\ref{CRs-scale} Panel 1, we see that slope gets consistent only above the mass $8\times 10^{13}\; M_{\odot}$ and the value is $L_{CR} \propto M^{1.2}$ which has also been observed in other studies with cooling and feedback physics \citep{Pfrommer_2008MNRAS}. This scaling does not work for the lower masses. In the same figure, Panel 3, total velocity dispersion (DM plus baryon) against radius plot indicates that objects with virial radius 1 Mpc and above only show a power law relation with velocity dispersion or turbulent energy of the system, but smaller objects are not having any correlation with the velocity distribution.

\section{Numerical resolution study}\label{res-study}

Effect of numerical resolution on computed physical parameters is a crucial aspect to be considered in simulation studies. Though, the resolution of our simulations is not among the highest one's, our resolution tests show it is adequate for our study. The main runs that are used for this study are performed with the cosmological and simulation parameters described in section~\ref{sample} with 6 levels of total (uni-grid + AMR) refinement leading to a resolution of $\sim$30 kpc. Keeping all other parameters same, we have further simulated some of our objects with different AMR levels to achieve different levels of resolution and compared the physical parameters obtained from them. With 5, 6 and 7 levels of total refinement, we have reached upto $\sim$ 60, 30 and 15 kpc at the highest resolution level and used these three resolutions for testing the convergence of the results.

\begin{figure*}
\includegraphics[width=6cm,trim={0.1cm 0.1cm 0.2cm 0.3cm},clip]{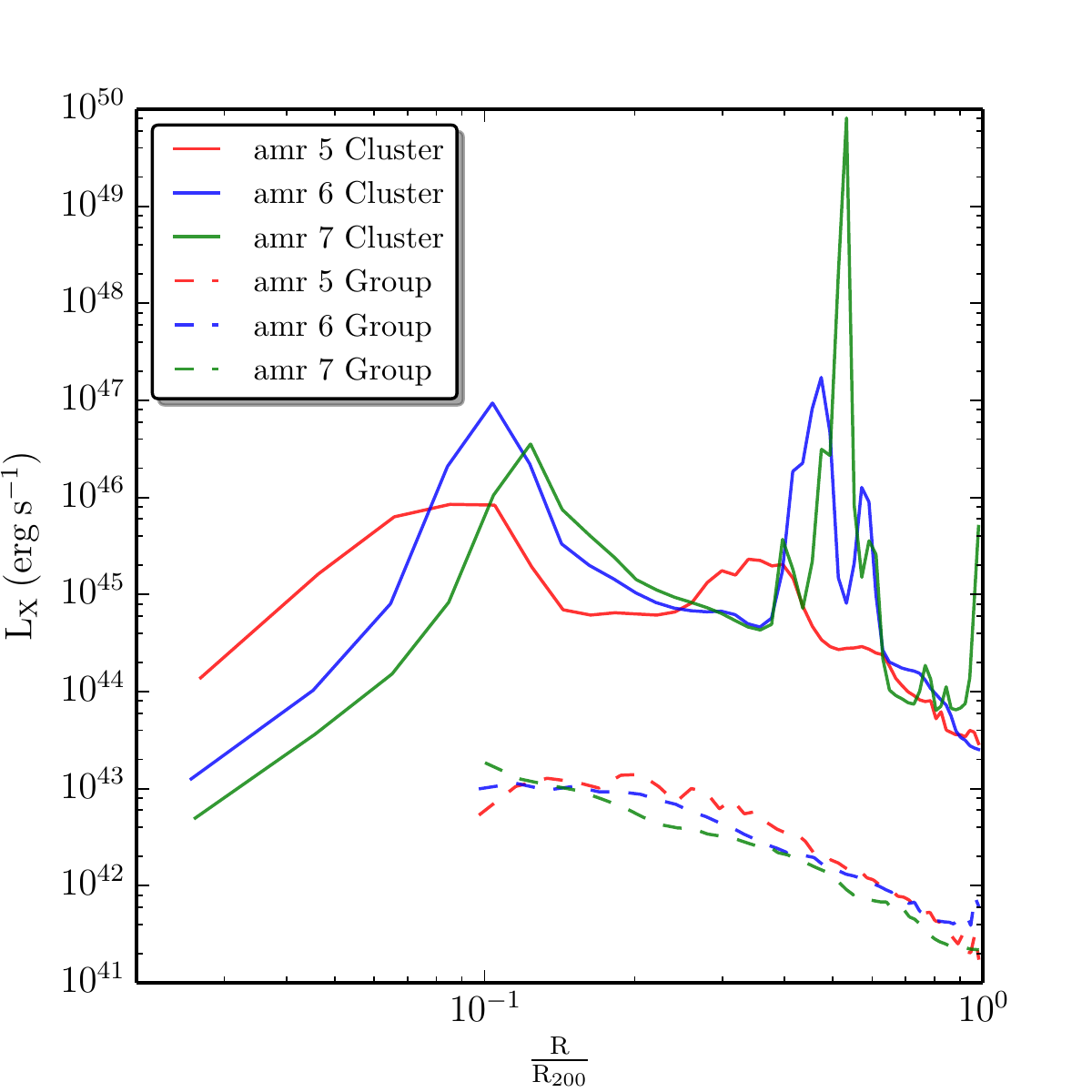}\hspace{-0.4cm}
\includegraphics[width=6cm,trim={0.1cm 0.1cm 0.2cm 0.3cm},clip]{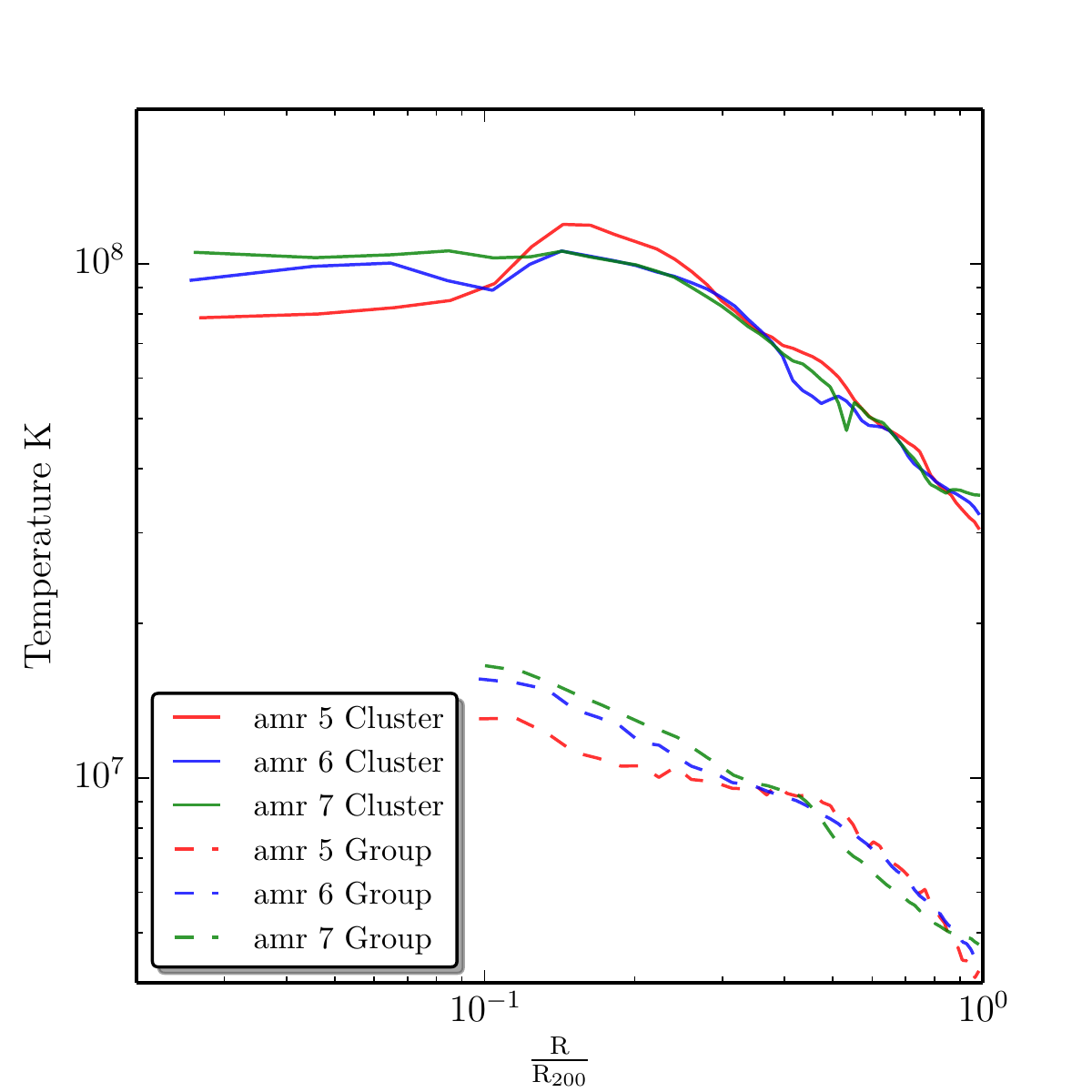}\hspace{-0.4cm}
\includegraphics[width=6cm,trim={0.1cm 0.1cm 0.2cm 0.3cm},clip]{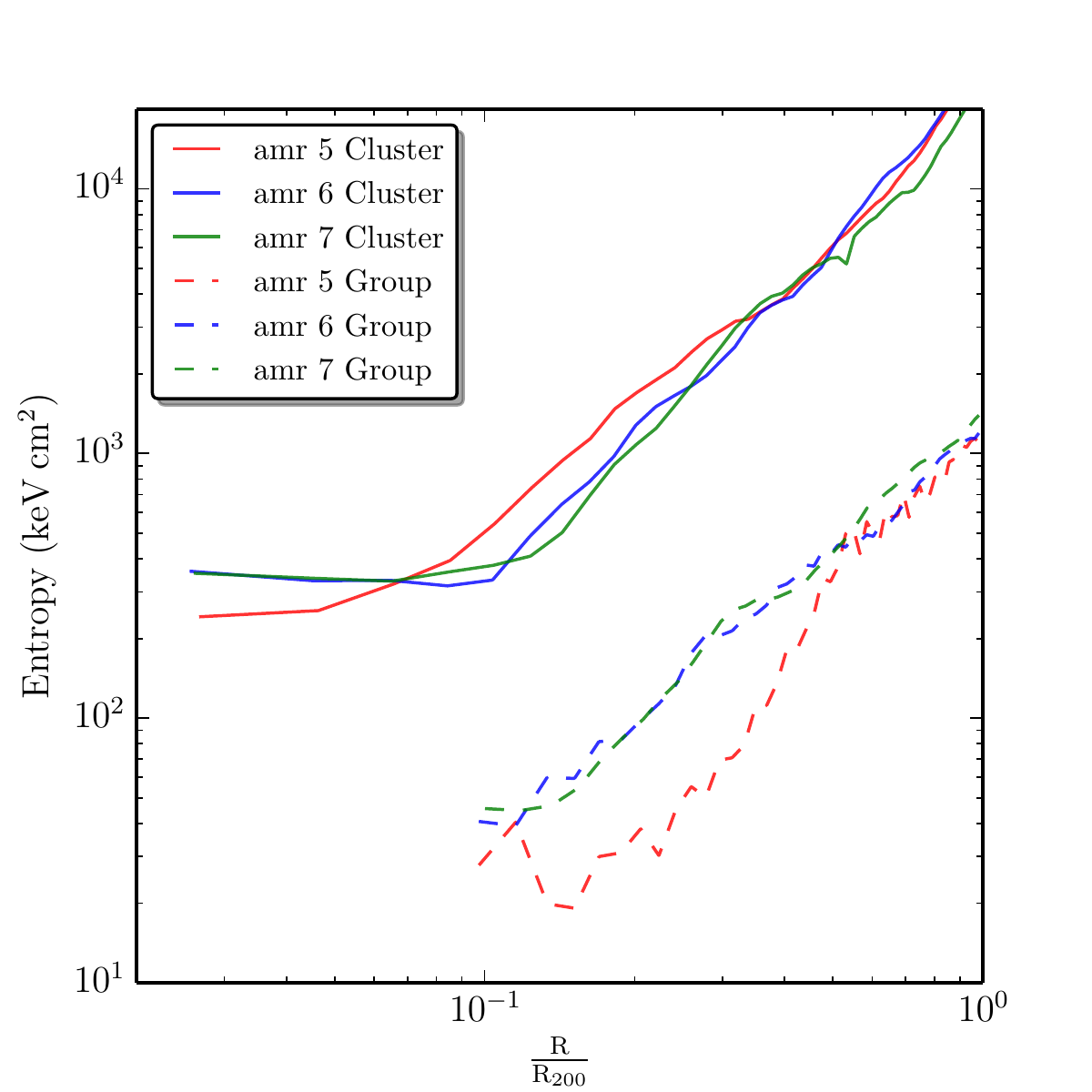}\hspace{-0.4cm}
\caption{X-ray luminosity, temperature and entropy have been plotted against normalized radius (normalized to $\rm{r_{200}}$) in the Panel 1,2 \& 3 respectively for a galaxy cluster ($\sim$ 10$^{15} \rm{M_{\odot}}$) and a group ($\sim$ 10$^{13} \rm{M_{\odot}}$) and for three resolutions namely LOWRES, REFRES and HIGHRES (colours as indicated in the legend). }
\label{radial-res-stud1}
\end{figure*}

\begin{figure*}
\includegraphics[width=4.6cm,trim={0.1cm 0.1cm 0.2cm 0.3cm},clip]{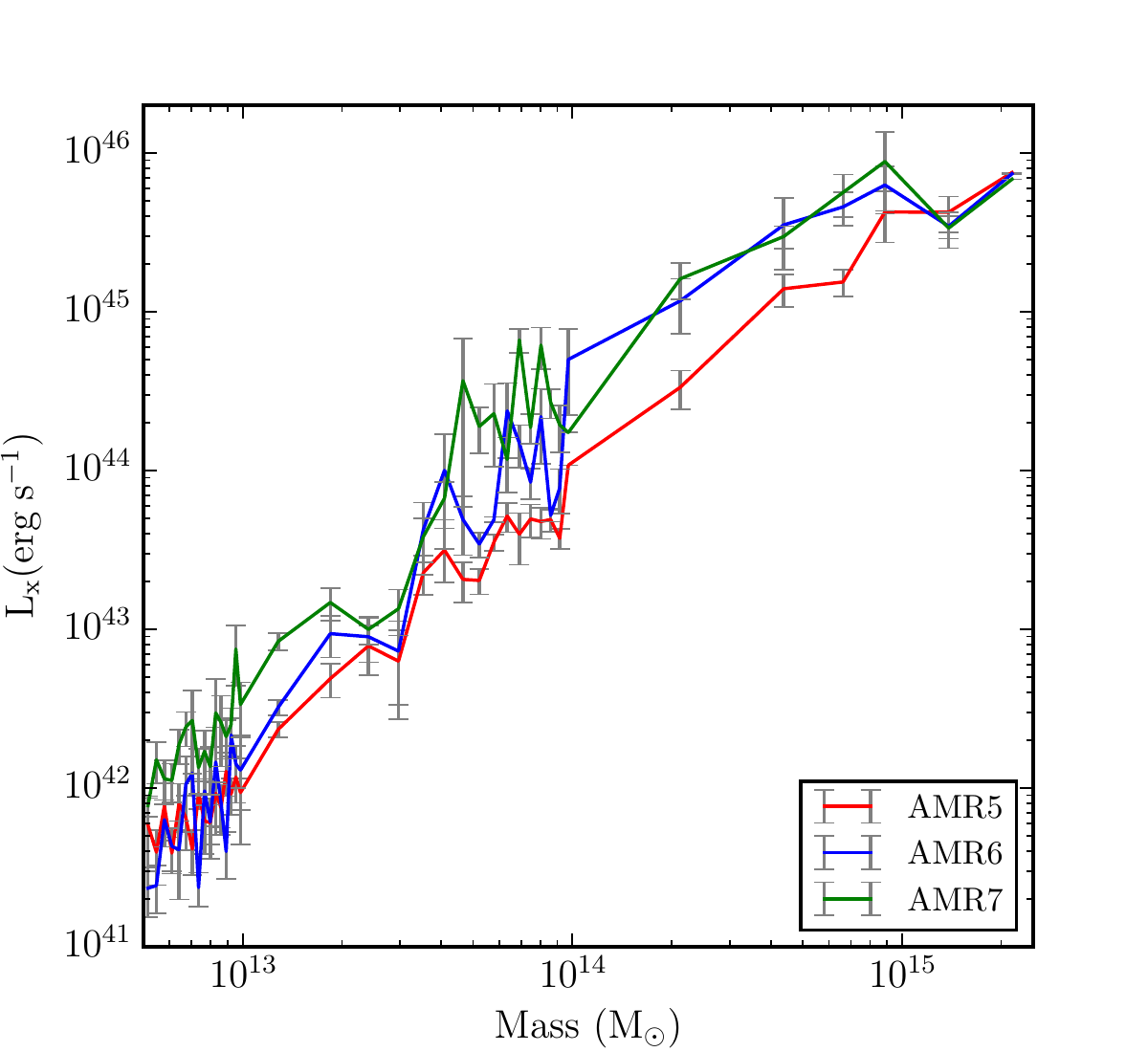}\hspace{-0.4cm}
\includegraphics[width=4.6cm,trim={0.1cm 0.1cm 0.2cm 0.3cm},clip]{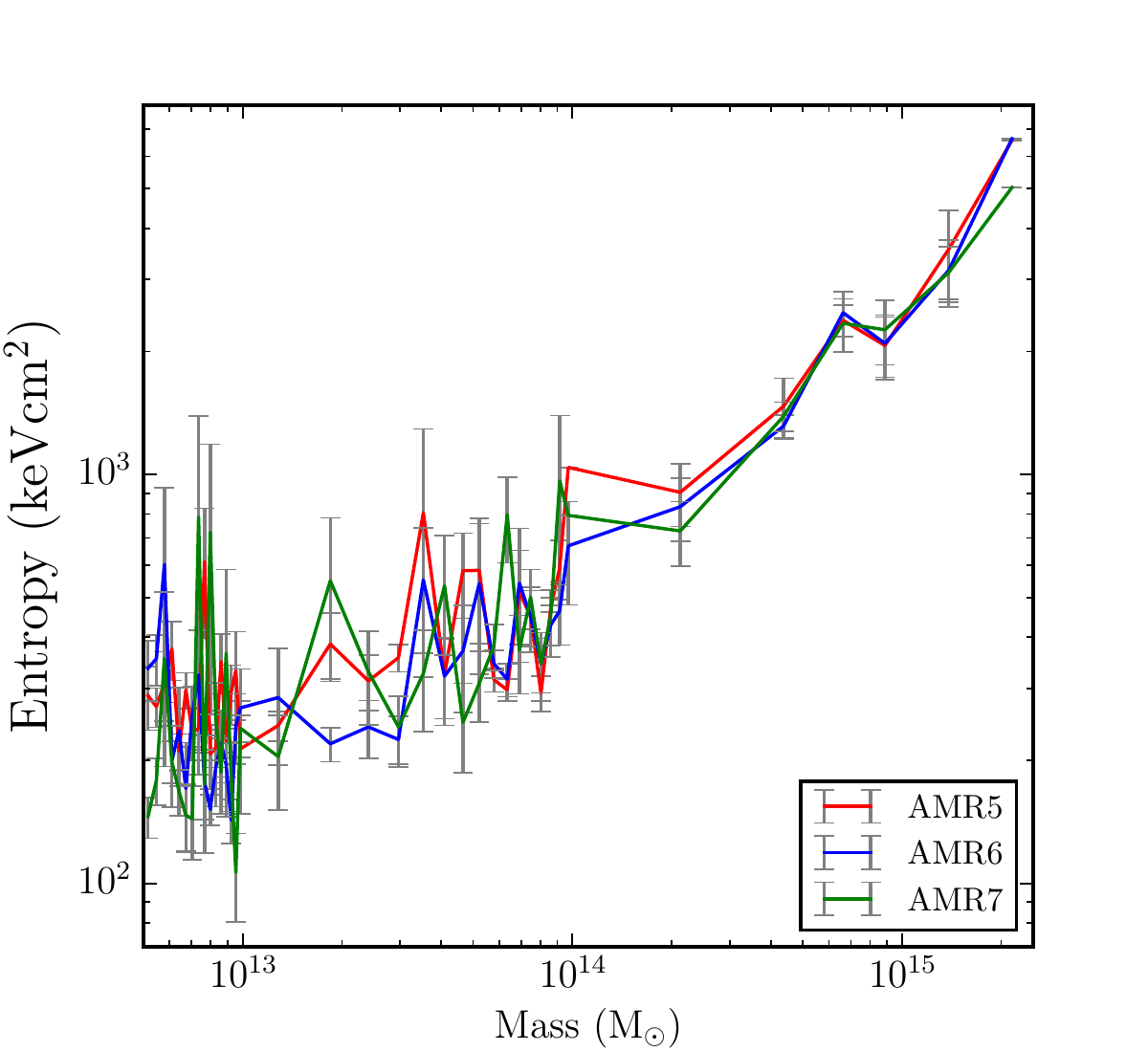}\hspace{-0.4cm}
\includegraphics[width=4.6cm,trim={0.1cm 0.1cm 0.2cm 0.3cm},clip]{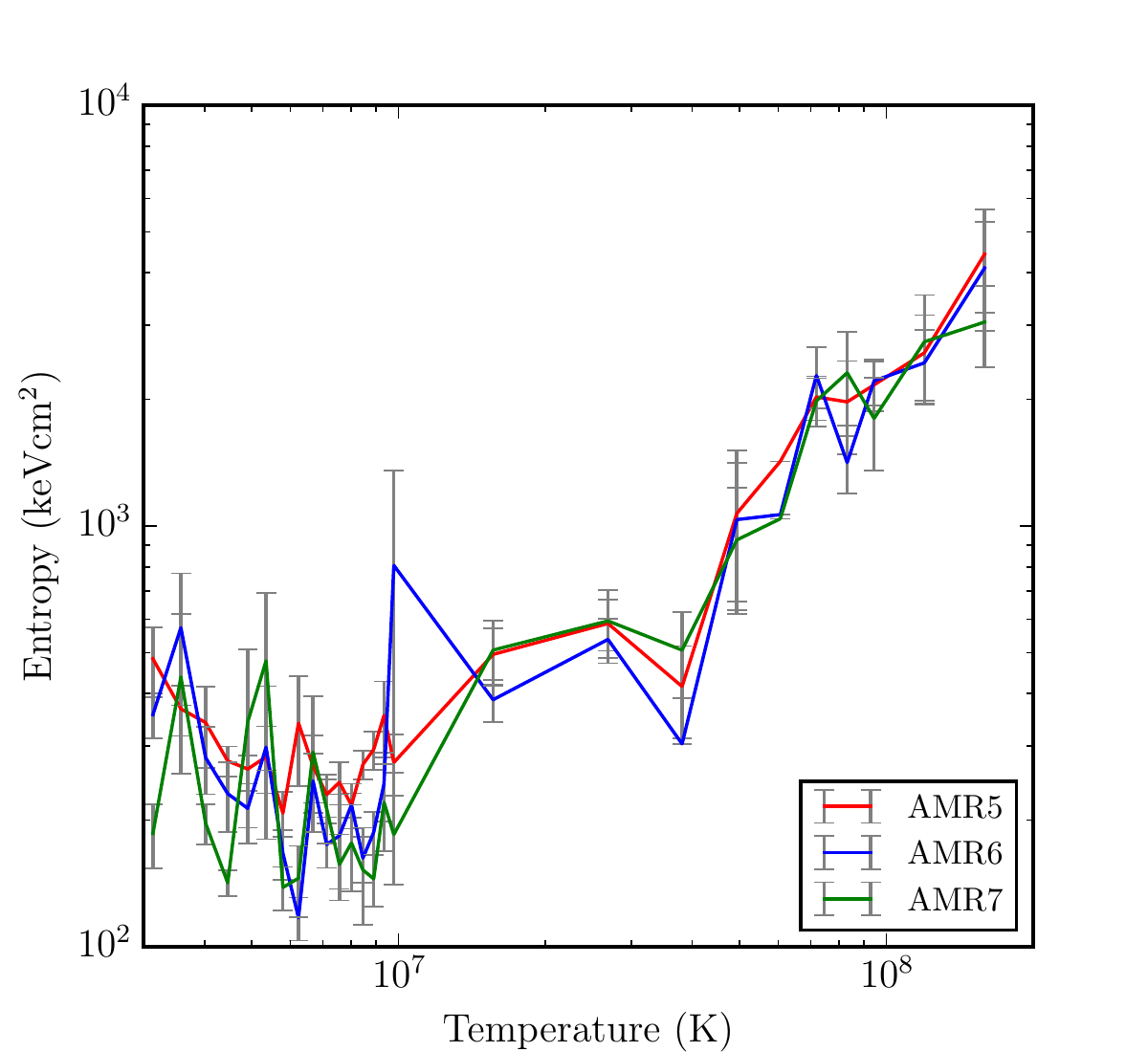}\hspace{-0.4cm}
\includegraphics[width=4.2cm,trim={0.1cm 0.1cm 0.2cm 0.3cm},clip]{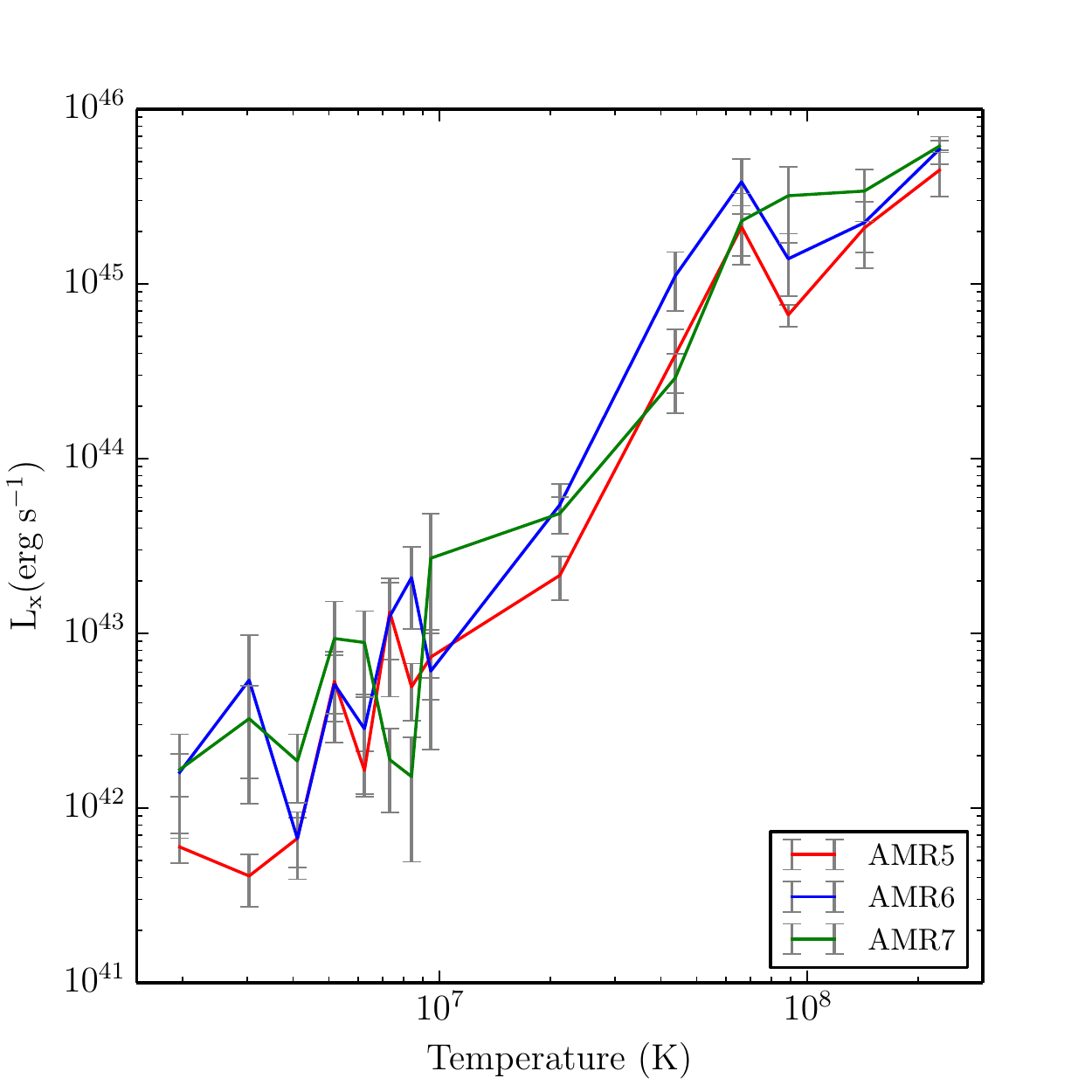}\hspace{-0.4cm}
\caption{Statistical average value and standard error of X-ray Luminosity and entropy from our sample data set has been plotted against virial mass ($\rm{M_{200}}$) in the Panel 1 \& 2 respectively for all three resolutions. Similarly, X-ray luminosity and entropy has been plotted against temperature in the Panel 3 \& 4 respectively for all three resolutions.}
\label{stat-res-stud2}
\end{figure*}

In fig~\ref{radial-res-stud1}, we have plotted radial variations of different physical parameters for a group  ($\sim$ 10$^{13} \rm{M_{\odot}}$) and a cluster ($\sim$ 10$^{15} \rm{M_{\odot}}$) that are almost in relaxed phase. X-ray luminosity, gas temperature and entropy have been plotted against normalised virial radius for three different resolutions. It can be noticed that REFRES simulation is almost same as the HIGHRES resolution with very little deviation though, LOWRES data are little away. In panel 1, the X-ray luminosity has a little spatial variation that usually occurring due to resolution sensitivity of transient phenomena like the shocks. The overall value of the respective parameters though does not get effected much by the resolution. This results show that our simulated parameters are almost converging with resolution that we took as the reference set of simulations i.e. $\sim$30 kpc with 6 levels of refinement. For further confirmation, a general study has been done with our total sample set. In Fig~\ref{stat-res-stud2}, X-ray luminosity has been plotted against mass in wide range and at each point, standard error has been computed and plotted. It can be noticed that REFRES and HIGHRES data are significantly overlapping and mostly are within the error bars. We have also studied the break points for different resolutions and found almost no change in the break points as expected and discussed (See the Appendix, Figure~\ref{stat-res-stud4}~\&~\ref{stat-res-stud5}). Further, discussions about other parameters have been given in the Appendix~\ref{appen-res-study}. These studies confirm that resolution-wise our REFRES runs are adequate for our present work.

\section{Discussions}\label{discus}

Our simulations with coolSF model has been compared with the available observed data (see Section~\ref{x-ray-obs}). It can be noticed that, previous works, done with SPH codes and using similar physics as coolSF, could not match the observed data {to that extent  as ours} \citep{LeBrun_2014MNRAS,McCarthy2010MNRAS}. Better agreement of our data with the observed data can possibly be attributed to difference in reproduction of baryon fraction as well as entropy and better gas mixing in ENZO AMR (PPM) code over SPH code \citep{Hubber_2013MNRAS,Valdarnini_2011A&A,O'Shea_2005ApJS}. AMR schemes are also known for better shock capturing and appropriate shock heating. We have implemented shock refinement very carefully to resolve the whole cluster volume as explained in Section~\ref{sample}. All these factors may have contributed in production of apparently different results from coolSF in ENZO compared to SPH. So, with the chosen simulation parameters for this study, ENZO-AMR code has been able to adequately match the observations without AGN feedback model. Nevertheless, AGN feedback is important for correct core entropy computation and inclusion of AGN feedback and other additional physics may effect these results and thus becomes an interesting topic for future research.
 
Further, we have computed  $L_X$ - T, $\rm{L_X}$ -M, S-T scaling laws for simulated LSS with mass spanning from $5\times10^{12}\; \rm{M_{\odot}}$ to 2.5$\times 10^{15}\; \rm{M_{\odot}}$. Scaling of total energy, baryonic fraction and non-thermal cosmic ray luminosity with mass have also been worked out. All these studies are consistently indicating a breaking away point at about $8 \times 10^{13} \; \rm{M_{\odot}}$. This has enabled us to conclude that the LSS consists of two populations, one is above and another is below this breaking point. Henceforth, we will call objects with mass less than $8\times 10^{13}\;\rm{M_{\odot}}$ as `Galaxy groups' or `galaxy cabal' and with higher mass as 'Galaxy clusters'

\begin{figure*}
\includegraphics[width=6cm,trim={0.1cm 0.1cm 0.2cm 0.2cm},clip]{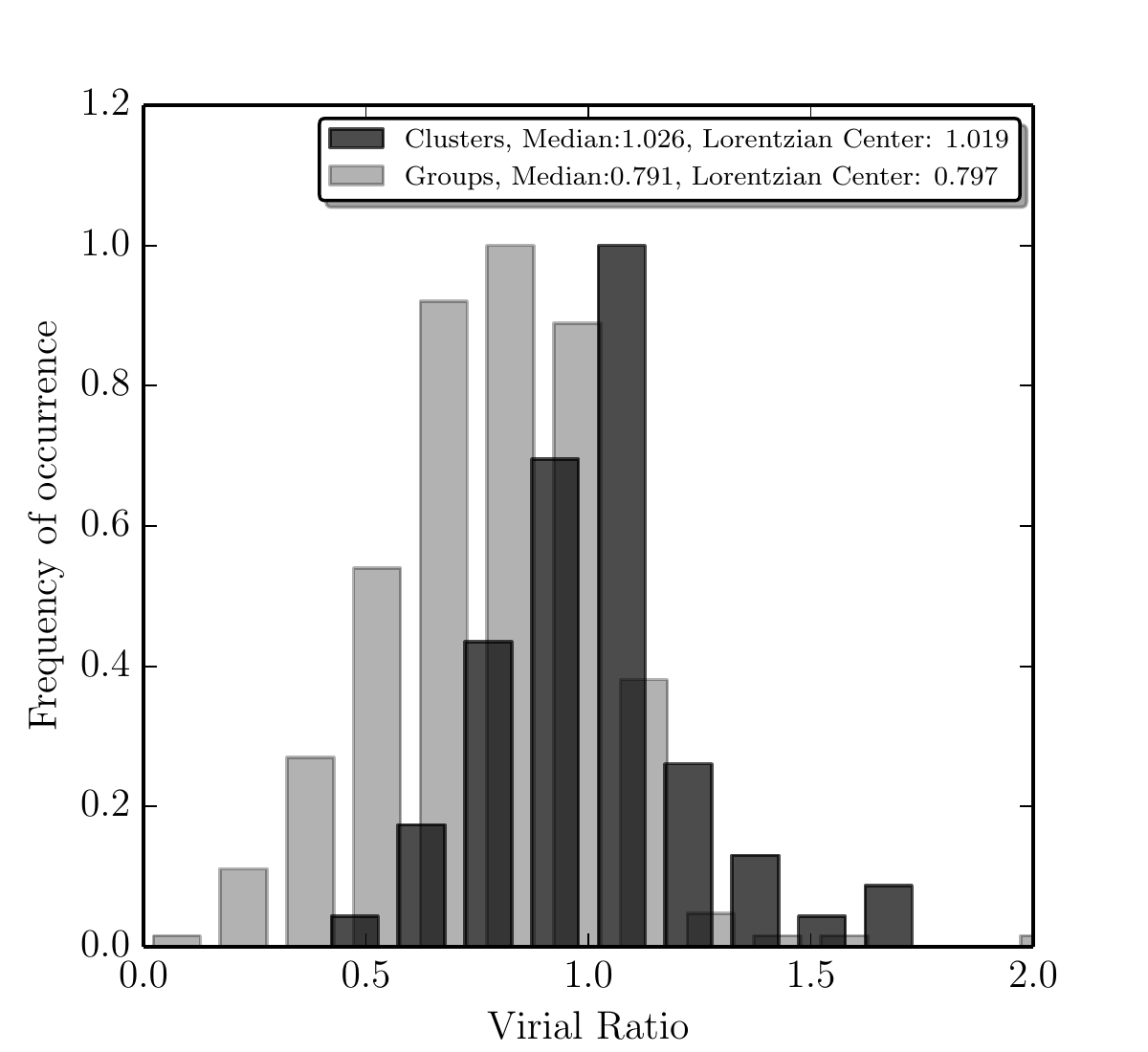}\hspace{-0.4cm}
\includegraphics[width=6cm,trim={0.1cm 0.1cm 0.2cm 0.2cm},clip]{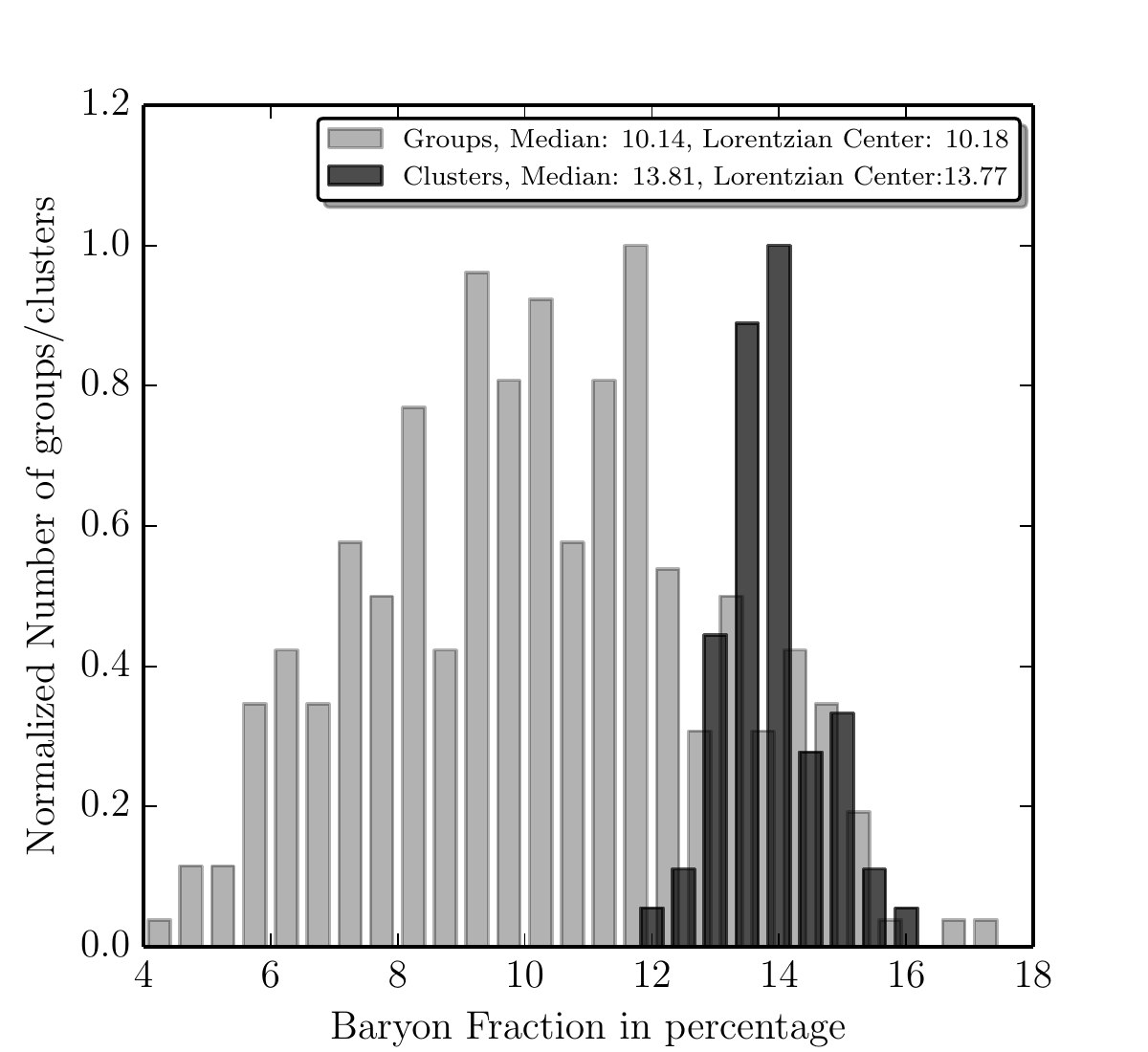}\hspace{-0.4cm}
\includegraphics[width=6cm,trim={0.1cm 0.1cm 0.2cm 0.2cm},clip]{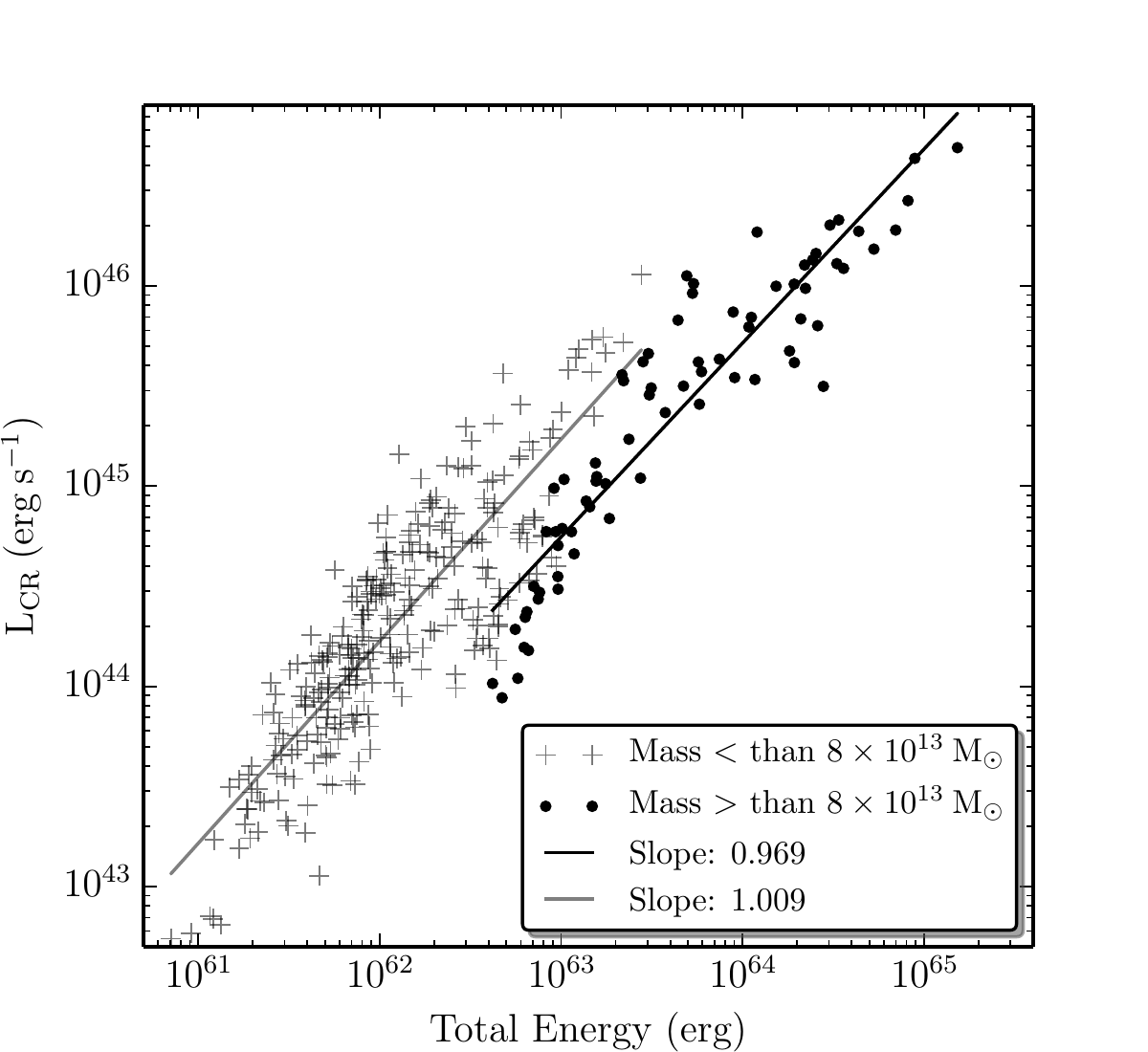}\hspace{-0.4cm}
\caption{In Panel 1 \& 2, normalised histogram of frequency of occurrence have been plotted for virial ratio and baryon fraction for the `galaxy groups' (light gray) and `galaxy clusters' (black). Panel 3, shows the cosmic ray luminosity vs total energy plot fitted for `galaxy clusters' and `groups'.}
\label{virial-ratio}
\end{figure*}

Galaxy clusters are mostly found to be relaxed systems and considered to be virialized \citep{Sarazin2003PhPl}. For a self gravitating system like the clusters, virial ratio can be expressed as $\xi$ = $\frac{(U_{int} + U_{ext}-E_s)}{2*KE} $ \citep{Davis_2011MNRAS}. Where, $U_{int}$ is the potential energy of the studied system and $U_{ext}$ is due to the mass outside the system radius (here, $r_{200}$) but, whose tidal effect can be felt. $E_s$ is the surface pressure term and KE is the kinetic energy of the system. For a perfectly virialised object, the ratio $\xi$ should be unity. We have computed the virial ratio for all objects in our sample set and plotted two different normalised histograms for the `galaxy groups' and `galaxy clusters' in Figure~\ref{virial-ratio}, Panel~1. Median and statistical (Lorentzian) peak of the `galaxy clusters' came out to be very close to unity (1.026 \& 1.019 respectively), indicating perfect virialization for most of the clusters. But, the median and peak value of virial ratio for `galaxy groups' are 0.791 \& 0.797 respectively i.e. far away from unity. This shows, groups are unstable and in hydrostatic disequilibrium, unlike the clusters and strongly supports the break point calculated from our study. 

Further, we have plotted baryon fraction histogram in Figure~\ref{virial-ratio}, Panel~2, which clearly shows bimodal distribution with two Lorentzian peaks at $\sim 10.18\%$ and $\sim 13.77\%$ and distinctly separated median at 10.14 \&  13.81 for groups and clusters respectively, just like the virial ratio distribution.  We have also plotted the cosmic ray luminosity as a function of total kinetic energy of the systems. Groups and Clusters are observed to follow two different evolutionary track as they fall into two parallel fitting lines separated by at least an order in energy (Figure~\ref{virial-ratio} Panel 3). 

\section{Summary}\label{conclude}
This research studies most of the possible scaling laws in thermal and non-thermal energies for the large scale objects in the framework of implemented baryon physics (see Section~\ref{sample}). The main takeaway points from this study are as follows.

$\#$ We could define clear distinguishing parameters for classifying `galaxy groups' and `clusters' for the first time. Strikingly, we found that cluster self similarity scales applied to the structures deviates away below a particular break away point in mass at $\sim 8\times 10^{13}\; M_{\odot}$ in all the studied parameters such as X-ray luminosity, temperature, baryon fractions and even in non-thermal cosmic ray luminosity.

$\#$ We also report that cluster properties deviates at temperature $1.16\times10^7$ K i.e. $\sim$1 keV and at a radius of $\sim$1 Mpc. So, for the first time, we are able to give a strong characteristic numbers to separate `galaxy groups' from the clusters and this study presents `galaxy groups' as a unique object in the structural hierarchy.

$\#$ CRs luminosity slope shown to be flatter in groups, indicating more non-thermal energy in them. This study also shows a very high level velocity dispersion i.e. turbulence in galaxy groups making them a test bed for the study of magnetisation and non-thermal emissions.

$\#$ From the virialization study it is established that `galaxy groups' are far away from the virialization and any estimation of physical parameters based on virial theorem would certainly go wrong. 

$\#$ Baryon fraction of galaxy groups are very low, and have large fluctuations in the values among the groups making them very unstable in nature. This opens up the door for research on modelling galaxy groups differently.




\section*{Acknowledgements}

This project is funded by DST INSPIRE Faculty (IFA-12/PH-44) and DST-SERB Fast Track scheme for young scientists, Grant No. SR/FTP/PS-118/2011. We are thankful to the Inter-University Centre for Astronomy and Astrophysics (IUCAA) for providing the HPC facility. RSJ would like to thank IUCAA for providing the research facilities as a visiting student. Computations described in this work were performed using the publicly-available \texttt{Enzo} code (http://enzo-project.org) and data analysis is done with the yt-tools (http://yt-project.org/). Authors would also like to thank the editor and the anonymous referee for their critical comments that has definitely helped in improving the content of this paper.

\appendix 

\section{Resolution study}\label{appen-res-study}

\begin{figure*}
\includegraphics[width=4.6cm,trim={0.1cm 0.1cm 0.2cm 0.3cm},clip]{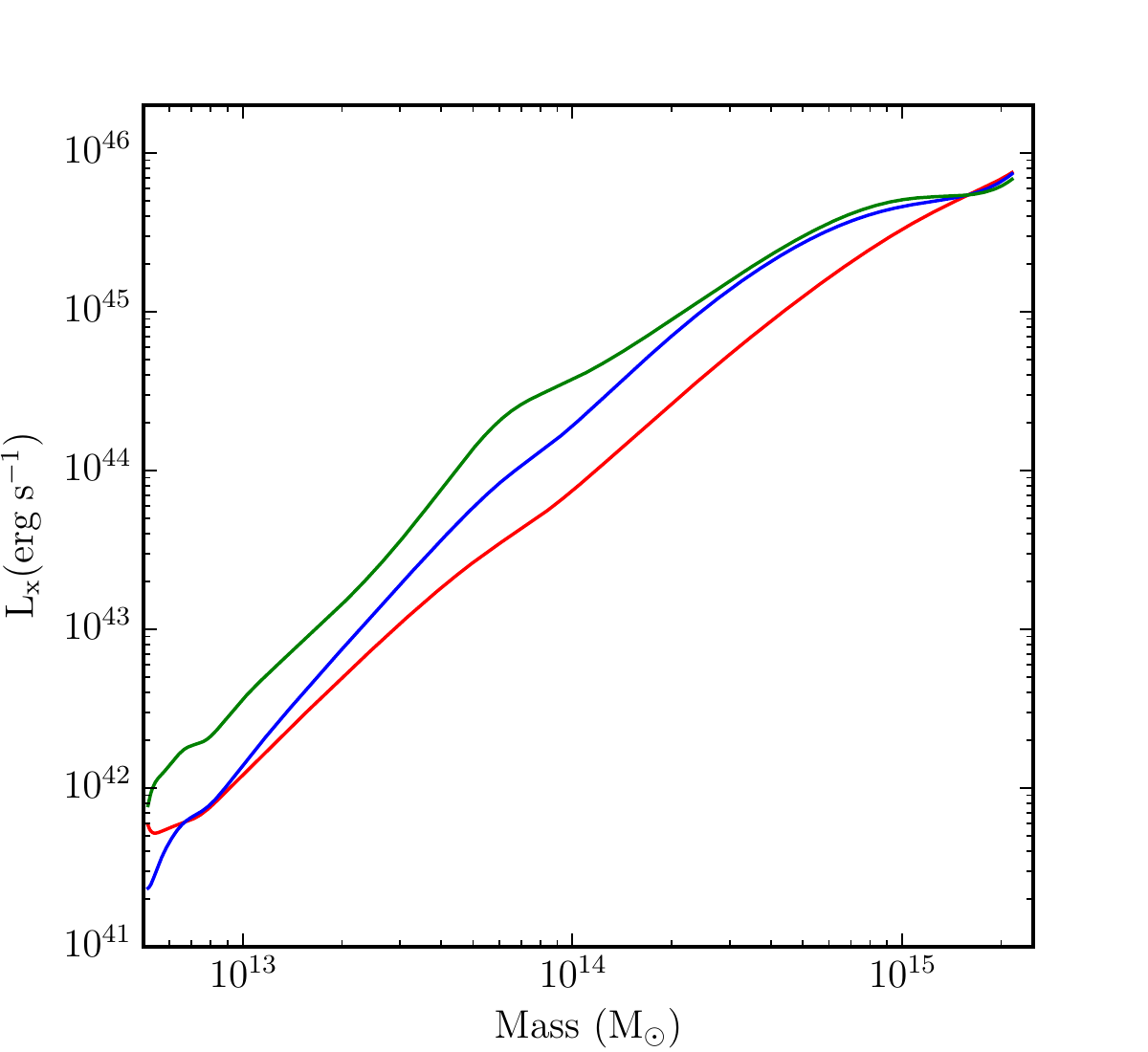}\hspace{-0.4cm}
\includegraphics[width=4.6cm,trim={0.1cm 0.1cm 0.2cm 0.3cm},clip]{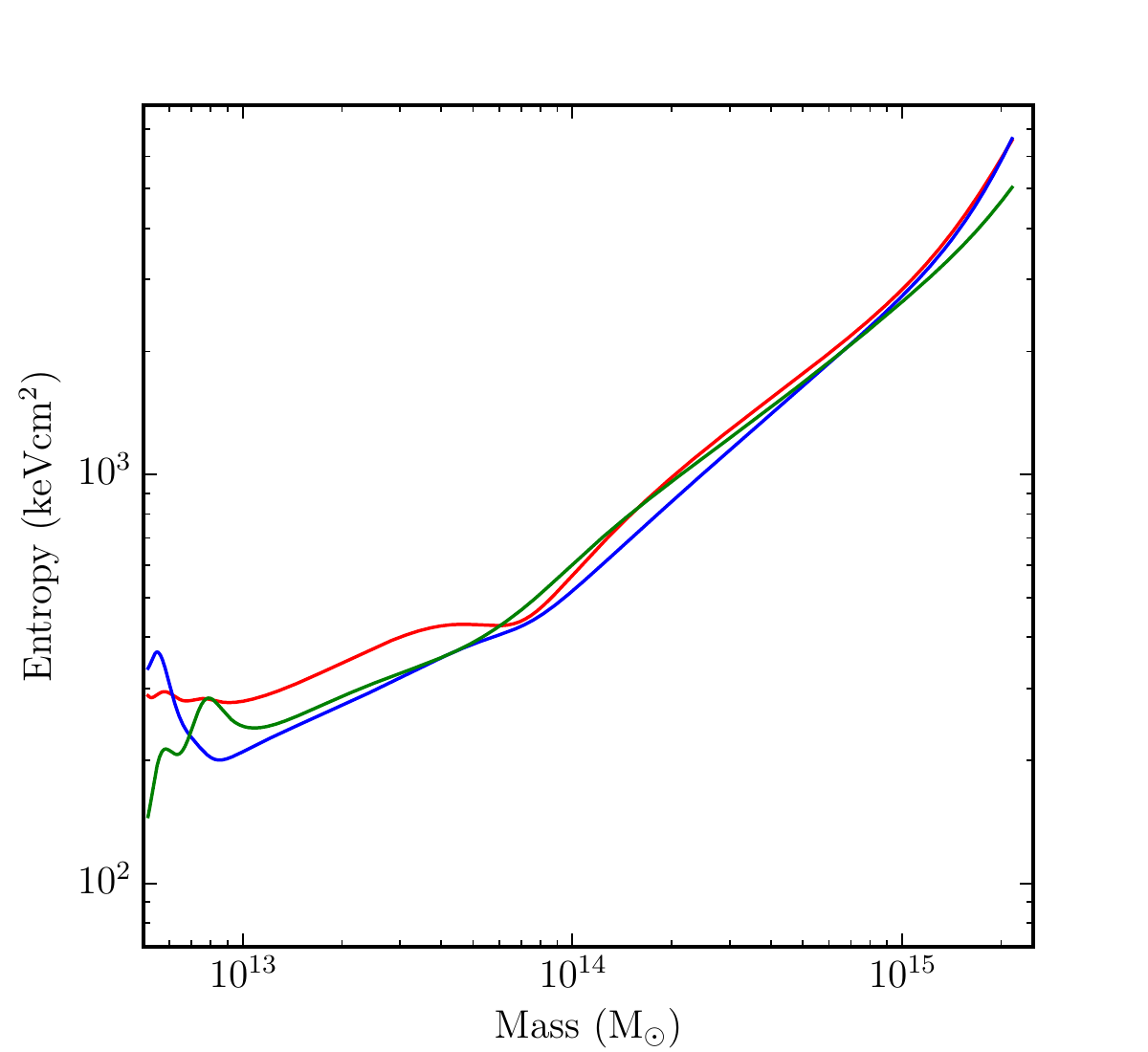}\hspace{-0.4cm}
\includegraphics[width=4.6cm,trim={0.1cm 0.1cm 0.2cm 0.3cm},clip]{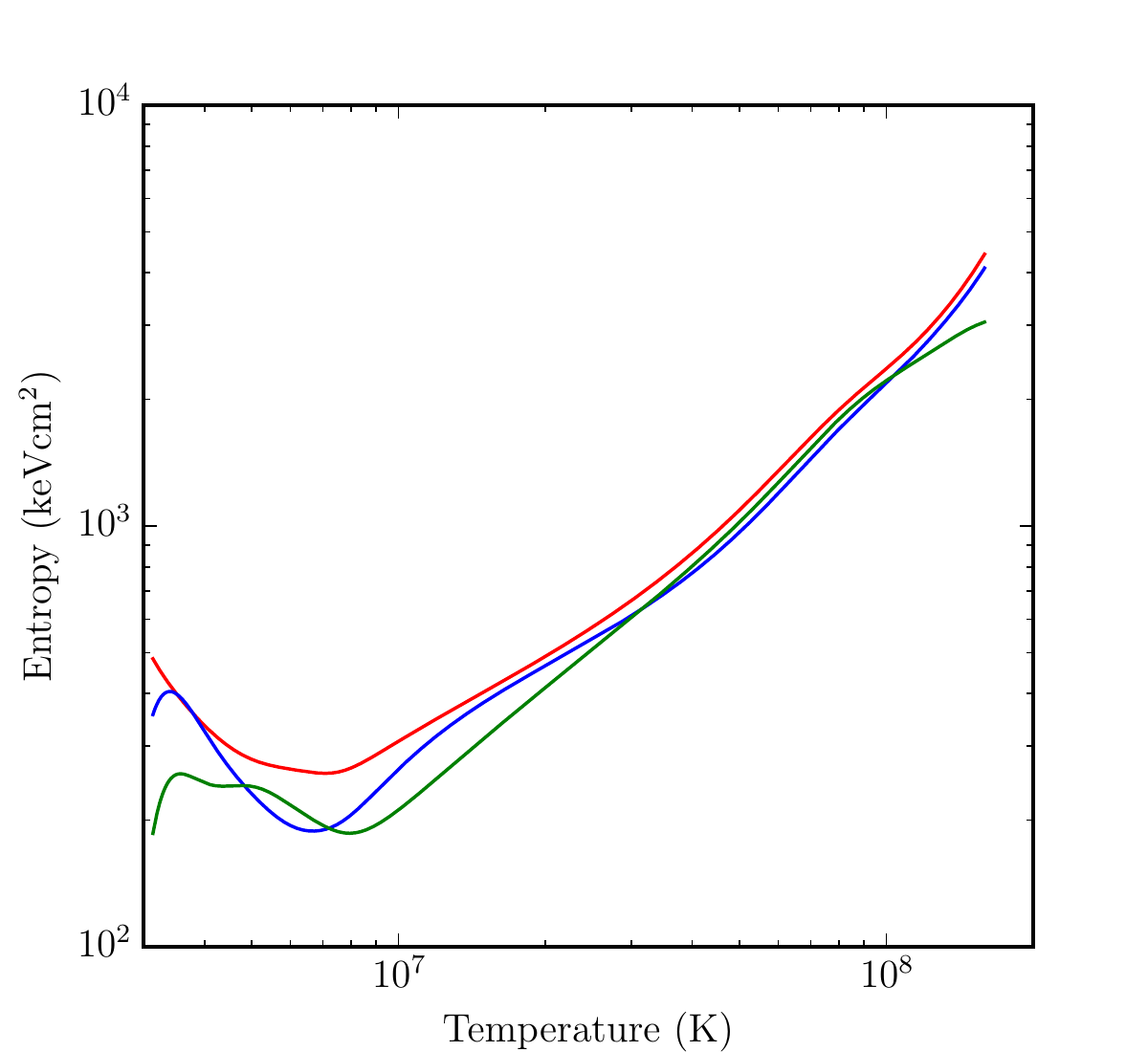}\hspace{-0.4cm}
\includegraphics[width=4.7cm,trim={0.1cm 0.1cm 0.2cm 0.3cm},clip]{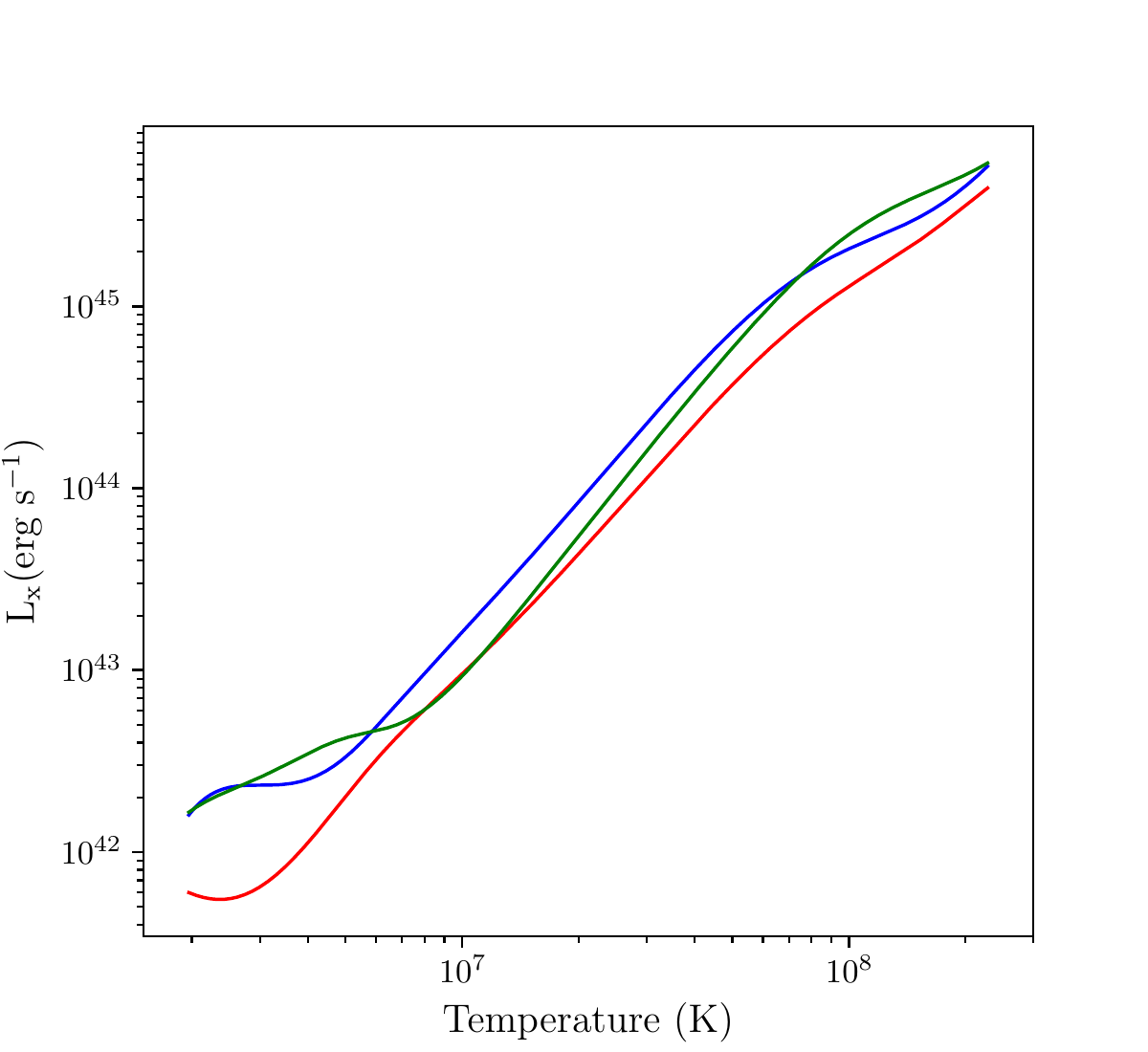}\hspace{-0.4cm}
\caption{For all three resolutions (LOWRES =`Red'; REFRES =`Blue'; HIGHRES =`Green'), B\'ezier curve has been fitted to statistical average value of X-ray Luminosity and entropy from our sample data and plotted against virial mass ($\rm{M_{200}}$) in the Panel 1 \& 2 respectively. Similarly, B\'ezier curve for X-ray luminosity and entropy has been plotted against temperature in the Panel 3 \& 4 respectively.}
\label{stat-res-stud3}
\end{figure*}

\begin{figure*}
\includegraphics[width=6cm,trim={0.1cm 0.1cm 0.2cm 0.3cm},clip]{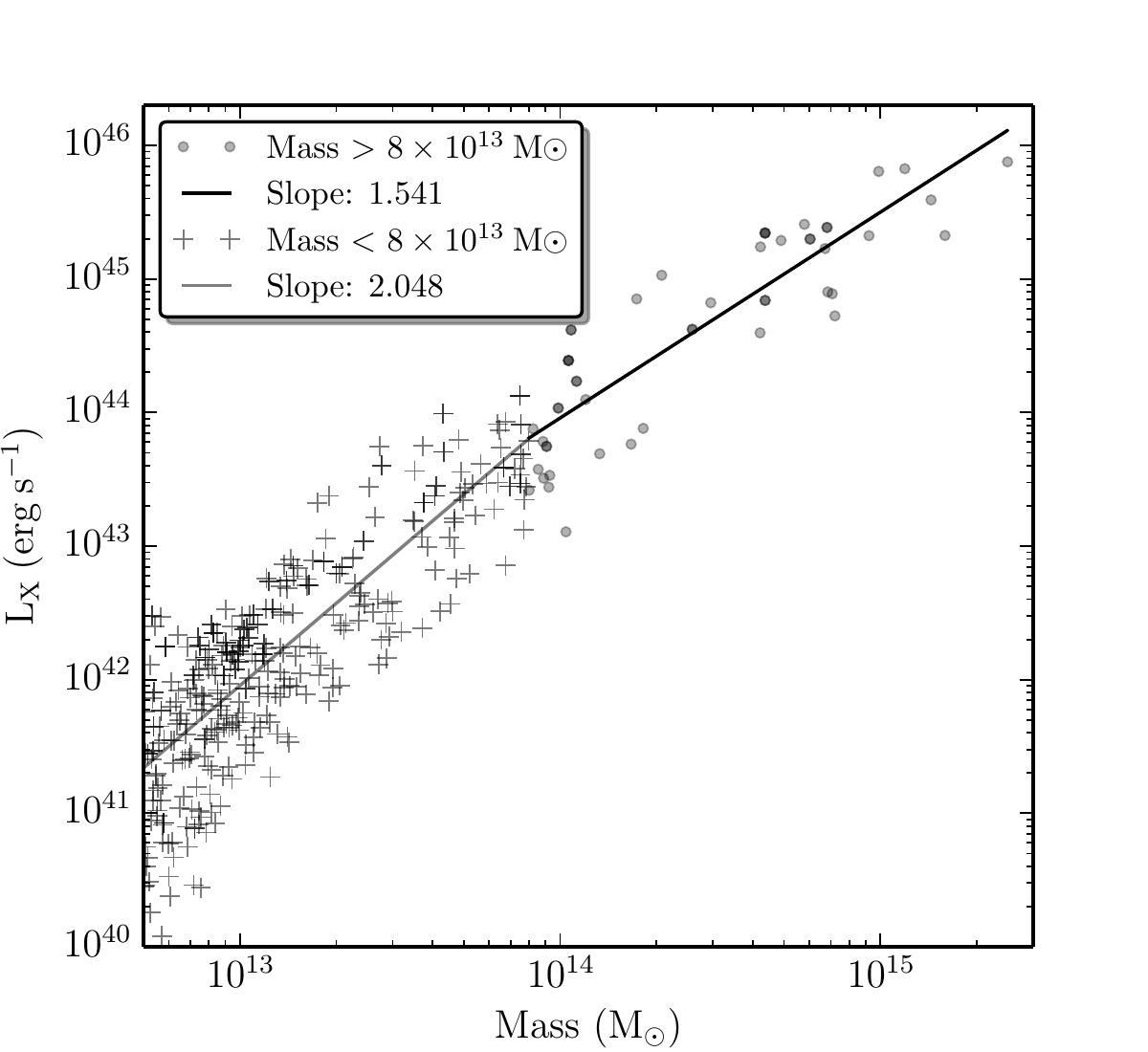}\hspace{-0.4cm}
\includegraphics[width=6cm,trim={0.1cm 0.1cm 0.2cm 0.3cm},clip]{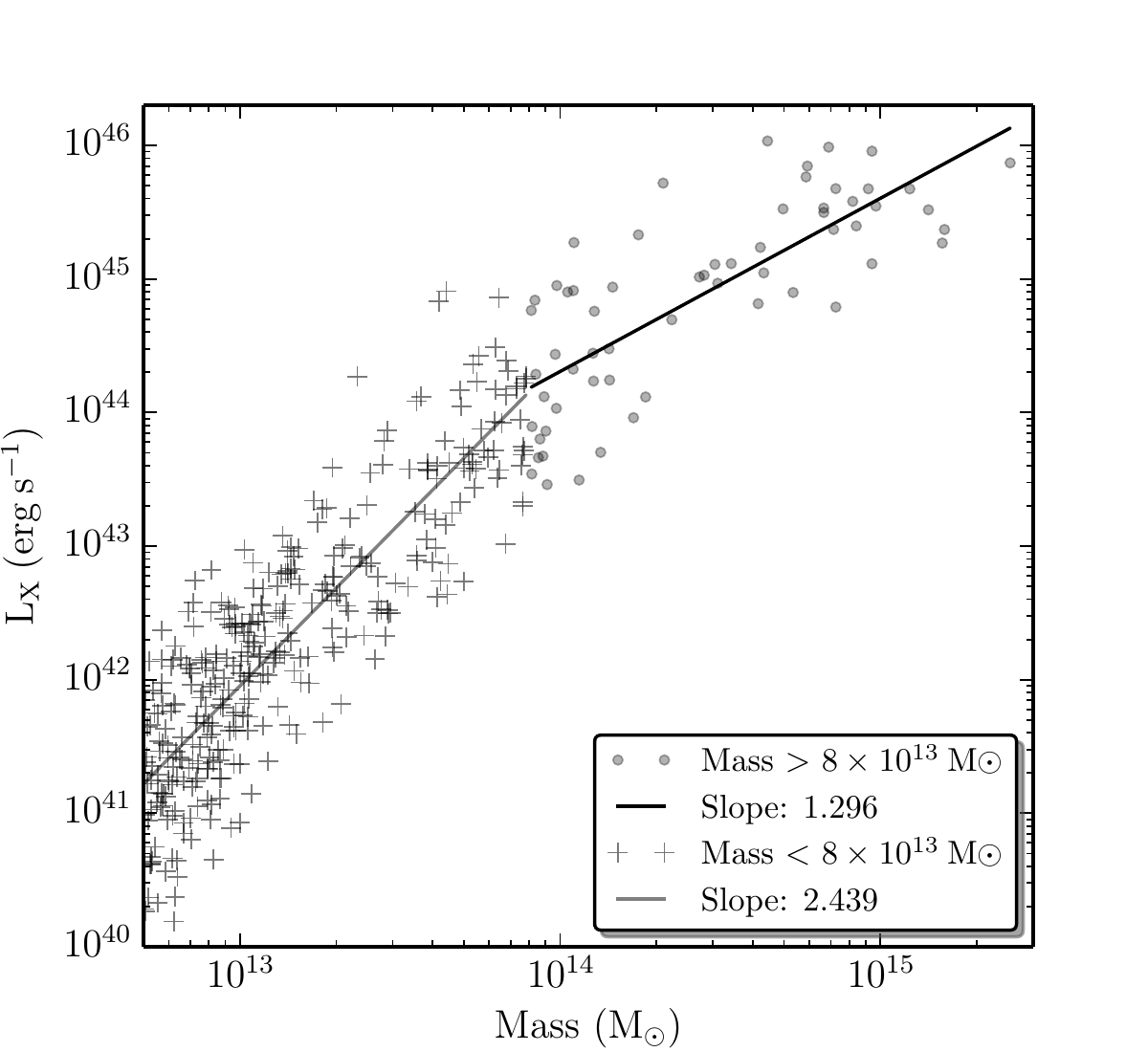}\hspace{-0.4cm}
\includegraphics[width=6cm,trim={0.1cm 0.1cm 0.2cm 0.3cm},clip]{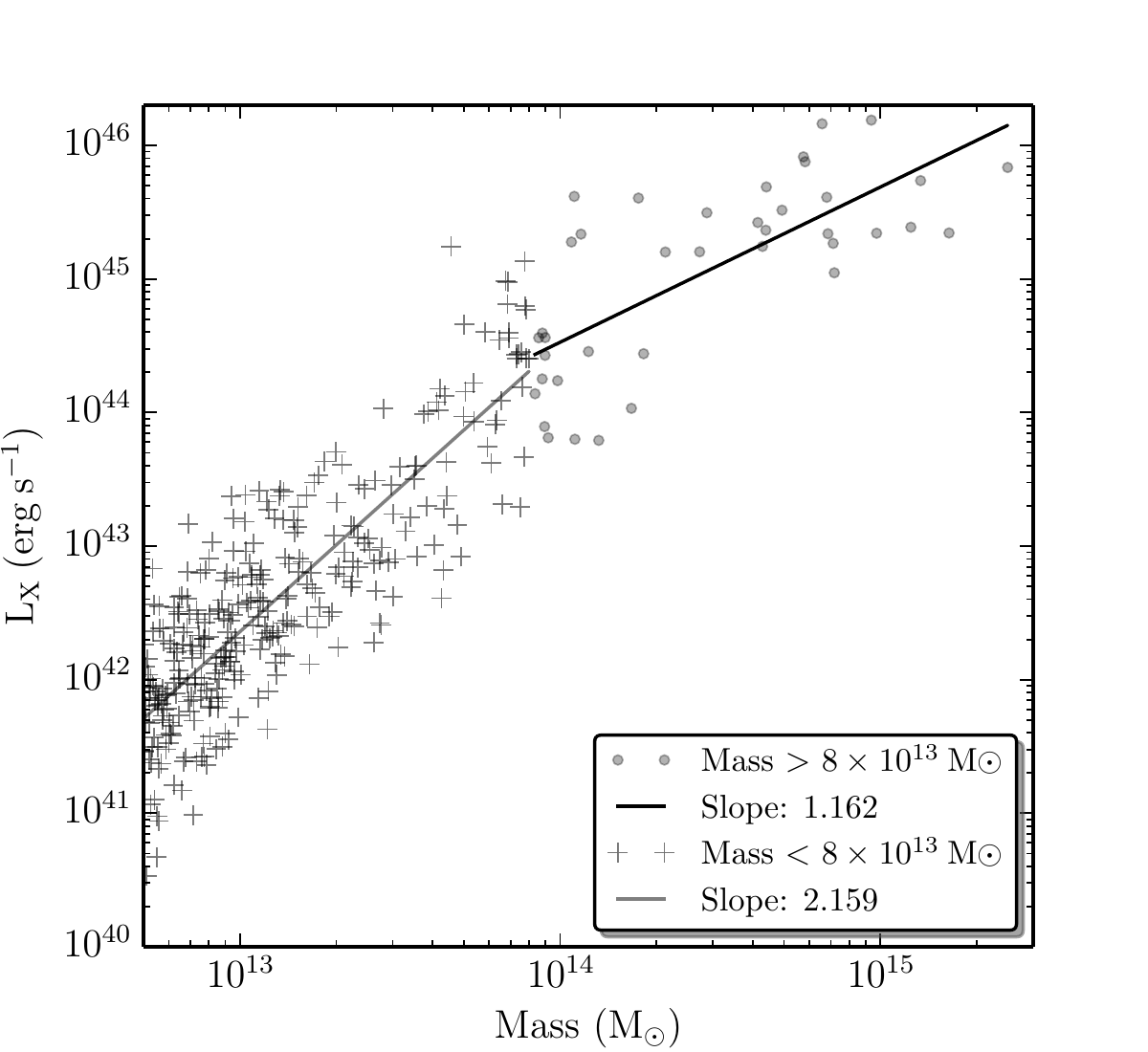}\hspace{-0.4cm}
\caption{X-ray Luminosity for LOWRES, REFRES and HIGHRES simulation data sets have been plotted against mass. In each cases, breaks in the fitted curves for clusters and groups are shown in the legend.}
\label{stat-res-stud4}
\end{figure*}

\begin{figure*}
\includegraphics[width=6cm,trim={0.1cm 0.1cm 0.2cm 0.3cm},clip]{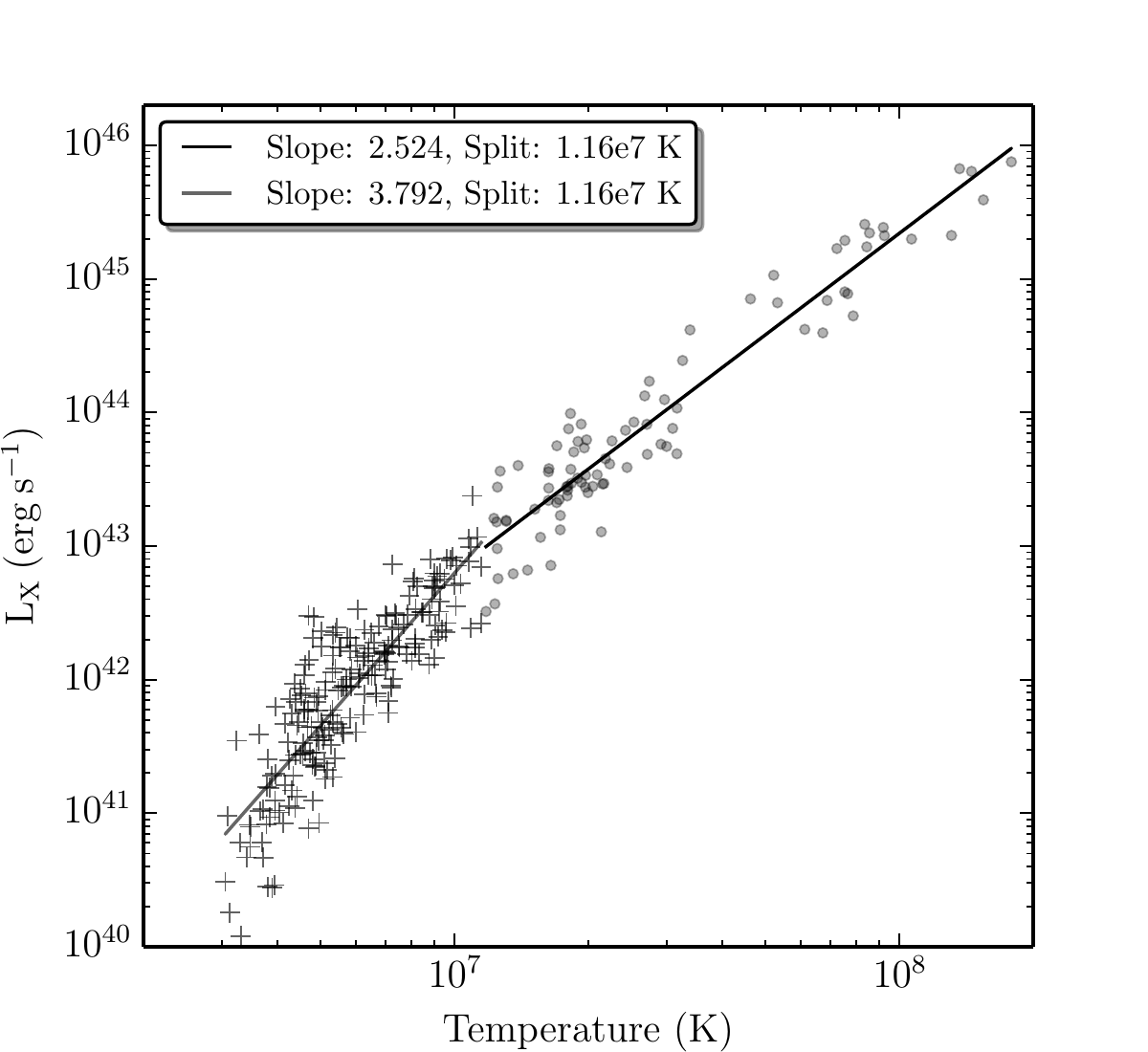}\hspace{-0.4cm}
\includegraphics[width=6cm,trim={0.1cm 0.1cm 0.2cm 0.3cm},clip]{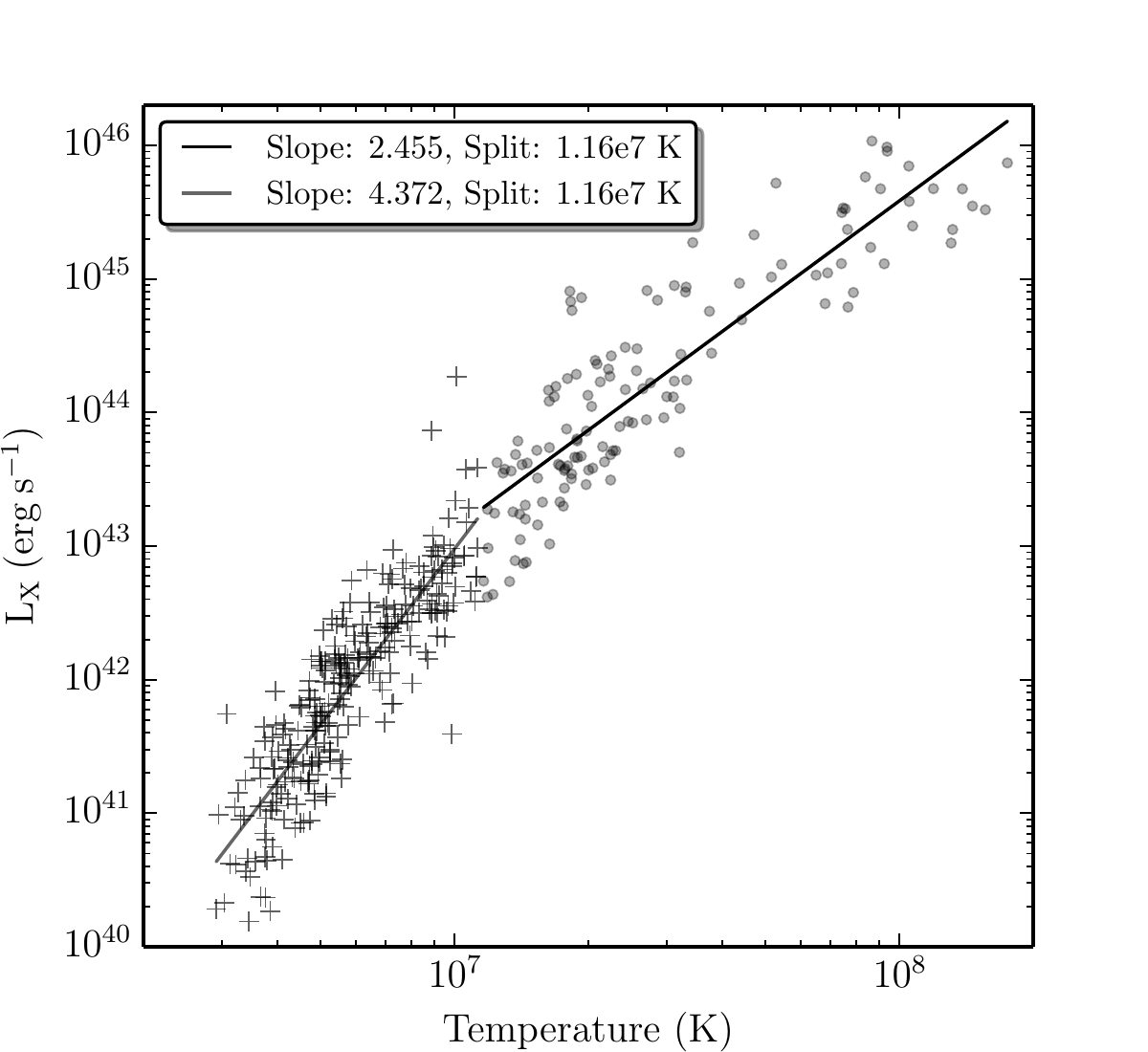}\hspace{-0.4cm}
\includegraphics[width=6cm,trim={0.1cm 0.1cm 0.2cm 0.3cm},clip]{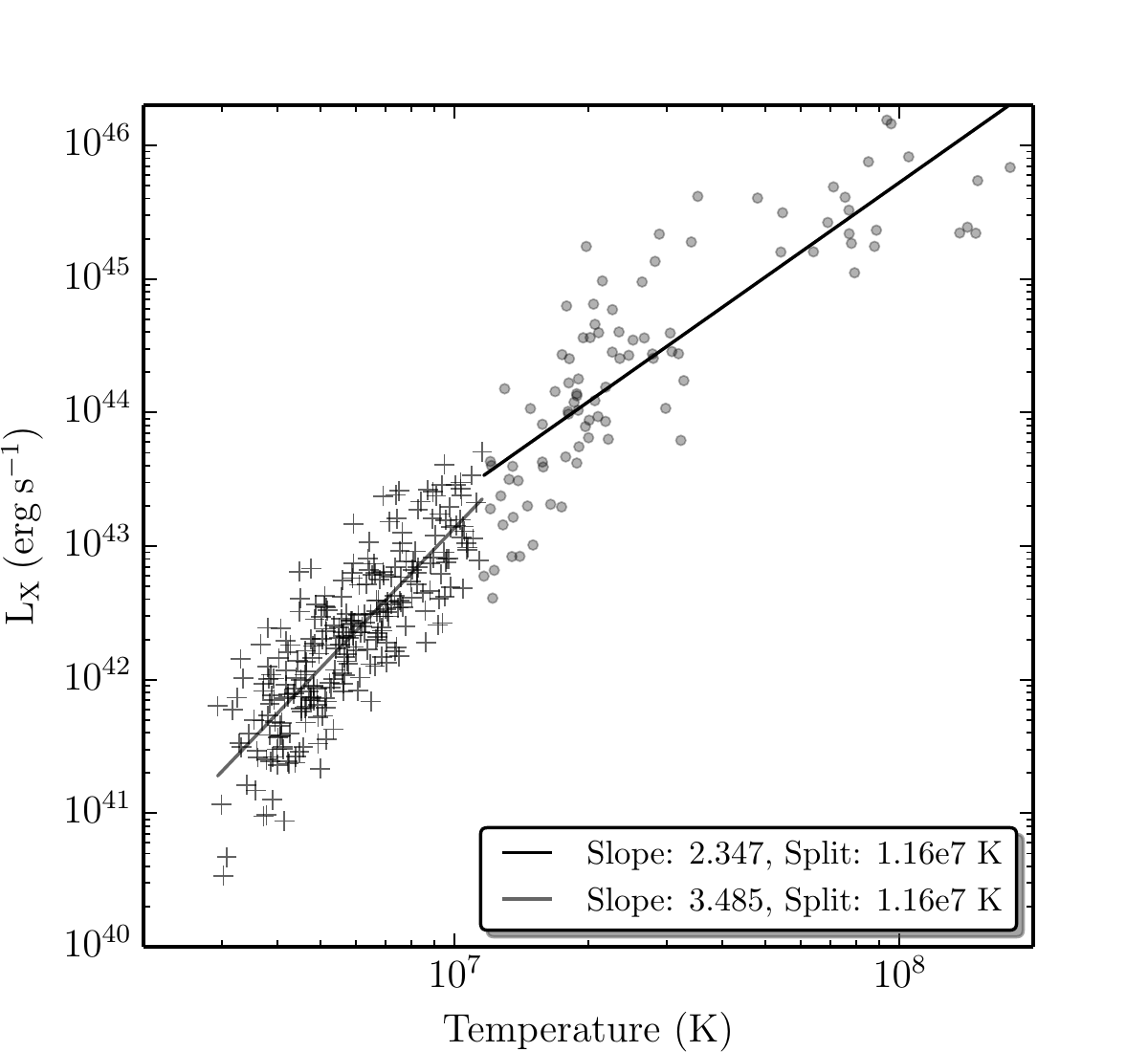}\hspace{-0.4cm}
\caption{Same as the Fig~\ref{stat-res-stud4} but, plotted against virial temperature.}
\label{stat-res-stud5}
\end{figure*}

Continuing from the Section~\ref{res-study} that deals with the resolution study, we have further studied a statistical trend in the Fig~\ref{stat-res-stud3} which shows a very good convergence of all the parameters especially for REFRES and HIGHRES simulations. Finally, breaks in scaling laws for Mass and Temperature has been plotted in Fig~\ref{stat-res-stud4}~\&~\ref{stat-res-stud5}. The break points in all the three resolutions came out to be very similar indicating a convergence of our results and confirms that there is a very little effect of resolution above our REFRES.

\bibliographystyle{mnras}

\begin{thebibliography}{99}
\bibitem[Allen \& Fabian(1998)]{Allen_1998MNRAS} 
Allen, S.~W., \& Fabian, A.~C.\ 1998, {\it MNRAS}, 297, L57
\bibitem[Andreon(2010)]{Andreon_2010MNRAS} 
Andreon, S.\ 2010, \mnras, 407, 263 
\bibitem[Arnaud \& Evrard(1999)]{Arnaud_1999MNRAS}
Arnaud, M., \& Evrard, A.~E.\ 1999, \mnras, 305, 631 
 \bibitem[Bharadwaj et al.(2015)]{Bharadwaj2015}
Bharadwaj, V., Reiprich, T.~H., Lovisari, L., \& Eckmiller, H.~J.\ 2015, {\it AAP}, 573, A75 
\bibitem[Bryan et al.(2014)]{Bryan_2014ApJS} 
Bryan, G.~L., Norman, M.~L., O'Shea, B.~W., et al.\ 2014, \apjs, 211, 19
\bibitem[Cen \& Ostriker(1992)]{Cen1992ApJ} 
Cen, R., \& Ostriker, J.~P.\ 1992, {\it ApJl}, 399, L113
\bibitem[Cole \& Lacey(1996)]{Cole_1996MNRAS} 
Cole, S., \& Lacey, C.\ 1996, \mnras, 281, 716
\bibitem[Costain et al.(1972)]{Costain_1972ApJ} 
Costain, C.~H., Bridle, A.~H., \& Feldman, P.~A.\ 1972, \apjl, 175, L15
\bibitem[Dai et al.(2010)]{Dai_2010ApJ} 
Dai, X., Bregman, J.~N., Kochanek, C.~S., \& Rasia, E.\ 2010, \apj, 719, 119
\bibitem[Dav{\'e} et al.(2002)]{Dave_2002ApJ} 
Dav{\'e}, R., Katz, N., \& Weinberg, D.~H.\ 2002, \apj, 579, 23
\bibitem[Davis et al.(2011)]{Davis_2011MNRAS} 
Davis, A.~J., D'Aloisio, A., \& Natarajan, P.\ 2011, {\it MNRAS}, 416, 242
\bibitem[Diaferio et al.(1993)]{Diaferio1993AJ} 
Diaferio, A., Ramella, M., Geller, M.~J., \& Ferrari, A.\ 1993, {\it AJ}, 105, 2035
\bibitem[D{\'{\i}}az-Gim{\'e}nez \& Mamon(2010)]{Diaz2010MNRAS} 
D{\'{\i}}az-Gim{\'e}nez, E., \& Mamon, G.~A.\ 2010, {\it MNRAS}, 409, 1227
\bibitem[Drury(1983)]{Drury1983RPPh} 
Drury, L.~O.\ 1983, Reports on Progress in Physics, 46, 973
\bibitem[Eisenstein \& Hu(1998)]{Eisenstein_1998ApJ} 
Eisenstein, D.~J., \& Hu, W.\ 1998, \apj, 496, 605 
\bibitem[Felten et al.(1966)]{Felten_1966ApJ} 
Felten, J.~E., Gould, R.~J., Stein, W.~A., \& Woolf, N.~J.\ 1966, \apj, 146, 955 
\bibitem[Ferland et al.(1998)]{Ferland_1998PASP} 
Ferland, G.~J., Korista, K.~T., Verner, D.~A., et al.\ 1998, \pasp, 110, 761
\bibitem[Gaspari et al.(2011)]{Gaspari2011MNRAS}
Gaspari, M., Brighenti, F., D'Ercole, A., \& Melioli, C.\ 2011,{\it MNRAS} , 415, 1549
\bibitem[Gilmour et al.(2007)]{Gilmour2007MNRAS} 
Gilmour, R., Gray,  M.~E., Almaini, O., et al.\ 2007, {\it MNRAS}, 380, 1467 
\bibitem[Giodini et al.(2009)]{Giodini_2009ApJ} 
Giodini, S., Pierini, D., Finoguenov, A., et al.\ 2009, \apj, 703, 982 
\bibitem[Girardi et al.(1996)]{Girardi_1996ApJ} 
Girardi, M., Fadda, D., Giuricin, G., et al.\ 1996, \apj, 457, 61
\bibitem[Gonzalez et al.(2013)]{Gonzalez_2013ApJ} 
Gonzalez, A.~H., Sivanandam, S., Zabludoff, A.~I., \& Zaritsky, D.\ 2013, \apj, 778, 14
\bibitem[Helsdon \& Ponman(2000)]{Helsdon_2000MNRAS} 
Helsdon, S.~F., \& Ponman, T.~J.\ 2000, \mnras, 315, 356
\bibitem[Henriksen \& Mushotzky(1986)]{Henriksen_1986ApJ} 
Henriksen, M.~J., \& Mushotzky, R.~F.\ 1986, \apj, 302, 287
\bibitem[Hubber et al.(2013)]{Hubber_2013MNRAS} 
Hubber, D.~A., Falle, S.~A.~E.~G., \& Goodwin, S.~P.\ 2013, \mnras, 432, 711
\bibitem[Jubelgas et al.(2008)]{Jubelgas2008A&A} 
Jubelgas, M., Springel, V., En{\ss}lin, T., \& Pfrommer, C.\ 2008,  {\it AAP} 481, 33
\bibitem[Kaiser(1986)]{kaiser1986MNRAS} 
Kaiser, N.\ 1986, \mnras, 222, 323
 \bibitem[Kang \& Ryu(2013)]{Kang_2013ApJ} 
Kang, H., \& Ryu, D.\ 2013, {\it ApJ}, 764, 95
\bibitem[Kitayama \& Suto(1996)]{Kitayama_1996ApJ} 
Kitayama, T., \& Suto, Y.\ 1996, \apj, 469, 480  
\bibitem[Komatsu et al.(2009)]{Komatsu_2009_APJS} 
Komatsu, E., Dunkley, J., Nolta, M.~R., et al.\ 2009, \apjs, 180, 330 
\bibitem[Komatsu \& Seljak(2002)]{Kamatsu2002MNRAS} 
Komatsu, E., \& Seljak, U.\ 2002, {\it MNRAS}, 336, 1256
\bibitem[Lagan{\'a} et al.(2011)]{Lagana_2011ApJ} 
Lagan{\'a}, T.~F., Zhang, Y.-Y., Reiprich, T.~H., \& Schneider, P.\ 2011, \apj, 743, 13
\bibitem[Lagan{\'a} et al.(2013)]{Lagana_2013A&A} 
Lagan{\'a}, T.~F., Martinet, N., Durret, F., et al.\ 2013, \aap, 555, A66 
\bibitem[Le Brun et al.(2014)]{LeBrun_2014MNRAS} 
Le Brun, A.~M.~C., McCarthy, I.~G., Schaye, J., \& Ponman, T.~J.\ 2014, \mnras, 441, 1270 
 \bibitem[Lietzen et al.(2012)]{lietzen2012}  
Lietzen, H., Tempel, E., Hein{\"a}m{\"a}ki, P., et al. 2012, {\it AAP}, 545, A104
\bibitem[Longair(1994)]{Longair1994} 
Longair, M.~S.\ 1994, High energy astrophysics.~Volume 2., by Longair, M.~S..~ Cambridge University Press, Cambridge (UK), 1994, 410 p., ISBN 0-521-43439-4
\bibitem[Lovisari et al.(2015)]{Lovisari2015} 
Lovisari, L., Reiprich, T.~H., \& Schellenberger, G.\ 2015, {\it AAP}, 573, A118
\bibitem[Markevitch(1998)]{Markevitch_1998ApJ} 
Markevitch, M.\ 1998, \apj, 504, 27
\bibitem[Mathis et al.(2005)]{Mathis_2005MNRAS} 
Mathis, H., Lavaux, G., Diego, J.~M., \& Silk, J.\ 2005, \mnras, 357, 801 
\bibitem[Maughan et al.(2012)]{Maughan_2012_MNRAS} 
Maughan, B.~J., Giles, P.~A., Randall, S.~W., Jones, C., \& Forman, W.~R.\ 2012, \mnras, 421, 1583 
\bibitem[McCarthy et al.(2010)]{McCarthy2010MNRAS} McCarthy, I.~G., Schaye, J., Ponman, T.~J., et al.\ 2010, \mnras, 406, 822
\bibitem[Miniati \& Beresnyak(2015)]{Miniati_2015Natur} 
Miniati, F., \& Beresnyak, A.\ 2015, \nat, 523, 59
\bibitem[Mittal et al.(2011)]{Mittal_2011A&A} 
Mittal, R., Hicks, A., Reiprich, T.~H., \& Jaritz, V.\ 2011, \aap, 532, A133
\bibitem[Morandi \& Sun(2016)]{morandi2016MNRAS} 
Morandi, A., \& Sun, M.\ 2016, \mnras, 457, 3266
\bibitem[Mulchaey(2000)]{Mulchaey2000ARA&A} 
Mulchaey, J.~S.\ 2000, \araa, 38, 289 
\bibitem[Navarro et al.(1996)]{Navarro_1996ApJ} 
Navarro, J.~F., Frenk, C.~S., \& White, S.~D.~M.\ 1996, \apj, 462, 563 
\bibitem[O'Shea et al.(2005)]{O'Shea_2005ApJS} 
O'Shea, B.~W., Nagamine, K., Springel, V., Hernquist, L., \& Norman, M.~L.\ 2005, \apjs, 160, 1
\bibitem[Osmond \& Ponman(2004)]{Osmond_2004_MNRAS} 
Osmond, J.~P.~F., \& Ponman, T.~J.\ 2004, \mnras, 350, 1511
\bibitem[Planelles et al.(2013)]{Planelles_2013MNRAS} 
Planelles, S., Borgani, S., Dolag, K., et al.\ 2013, \mnras, 431, 1487
\bibitem[Paul et al.(2011)]{Paul2011ApJ} 
Paul, S., Iapichino, L., Miniati, F., Bagchi, J., \& Mannheim, K.\ 2011, {\it ApJ}, 726, 17
\bibitem[Paul et al.(2015)]{Paul_2015fers.confE} 
Paul, S., Gupta, P., John, R.~S., \& Punjabi, V.\ 2015, The Many Facets of Extragalactic Radio Surveys: Towards New Scientific Challenges, 65 
\bibitem[Peebles(1980)]{Peebles_1980lssu.book} 
Peebles, P.~J.~E.\ 1980, Research supported by the National Science Foundation.~Princeton, N.J., Princeton University Press, 1980.~435 p.,
\bibitem[Pfrommer(2008)]{Pfrommer_2008MNRAS} 
Pfrommer, C.\ 2008, \mnras, 385, 1242 
\bibitem[Pratt et al.(2010)]{Pratt_2010A&A} 
Pratt, G.~W., Arnaud, M., Piffaretti, R., et al.\ 2010, \aap, 511, A85 
\bibitem[Rasia et al.(2014)]{Rasia_2014ApJ} 
Rasia, E., Lau, E.~T., Borgani, S., et al.\ 2014, \apj, 791, 96
\bibitem[Sanderson et al.(2013)]{Sanderson_2013MNRAS} 
Sanderson, A.~J.~R., O'Sullivan, E., Ponman, T.~J., et al.\ 2013, \mnras, 429, 3288
\bibitem[Sarazin \& White(1987)]{Sarazin1987ApJ} 
Sarazin, C.~L., \& White, R.~E., III 1987, {\it ApJ}, 320, 32 
\bibitem[Sarazin(2003)]{Sarazin2003PhPl} 
Sarazin, C.~L.\ 2003, Physics of Plasmas, 10, 1992 
\bibitem[Sarazin(2002)]{Sarazin_2002ASSL} 
Sarazin, C.~L.\ 2002, Merging Processes in Galaxy Clusters, 272, 1 
\bibitem[Sivakoff et al.(2008)]{Sivakoff2008ApJ} 
Sivakoff, G.~R., Martini, P., Zabludoff, A.~I., Kelson, D.~D., \& Mulchaey, J.~S.\ 2008, {\it ApJ}, 682, 803 
\bibitem[Skillman et al.(2008)]{Skillman_2008ApJ} 
Skillman, S.~W., O'Shea, B.~W., Hallman, E.~J., Burns, J.~O., \& Norman, M.~L.\ 2008, \apj, 689, 1063-1077 
\bibitem[Stott et al.(2012)]{Stott_2012MNRAS} 
Stott, J.~P., Hickox, R.~C., Edge, A.~C., et al.\ 2012, \mnras, 422, 2213
\bibitem[Sun et al.(2009)]{Sun_2009ApJ} 
Sun, M., Voit, G.~M., Donahue, M., et al.\ 2009, \apj, 693, 1142 
\bibitem[Sun(2012)]{Sun2012NJPh} 
Sun, M.\ 2012, New Journal of Physics, 14, 045004
\bibitem[Surajit et al.(2015)]{Surajit_2015fers.confE} 
Surajit, P., Gupta, P., John, R.~S., \& Punjabi, V.\ 2015, Proceedings of Science, 65, 65 
\bibitem[Tempel et al.(2014)]{tempel2014a} 
Tempel, E., Kipper, R., Saar, E., et al.\ 2014, {\it AAP}, 572, A8
\bibitem[Turk et al.(2011)]{Turk_2011ApJS} 
Turk, M.~J., Smith, B.~D., Oishi, J.~S., et al.\ 2011, {\it ApJS}, 192, 9
\bibitem[Valdarnini(2011)]{Valdarnini_2011A&A} 
Valdarnini, R.\ 2011, \aap, 526, A158
\bibitem[Vazza et al.(2011)]{Vazza_2011MNRAS} 
Vazza, F., Dolag, K., Ryu, D., et al.\ 2011, \mnras, 418, 960
\bibitem[Vikhlinin et al.(2006)]{Vikhlinin2006ApJ} 
Vikhlinin, A., Kravtsov, A., Forman, W., et al.\ 2006, \apj, 640, 691 
\bibitem[Voit et al.(2002)]{Voit_2002ApJ} 
Voit, G.~M., Bryan, G.~L., Balogh, M.~L., \& Bower, R.~G.\ 2002, \apj, 576, 601  
\bibitem[Xue \& Wu(2000)]{Xue_2000ApJ} 
Xue, Y.-J., \& Wu, X.-P.\ 2000, \apj, 538, 65
\bibitem[Zhang et al.(2011)]{Zhang_2011A&A} 
Zhang, Y.-Y., Lagan{\'a}, T.~F., Pierini, D., et al.\ 2011, \aap, 535, A78
\end{thebibliography}

%
%
%
%
%
%
%
\bsp	
\label{lastpage}
\end{document}